\documentclass[aps,prb,twocolumn,color,superscriptaddress,floatfix,nofootinbib]{revtex4-2}
\usepackage[english]{babel} 
\usepackage{amssymb}
\usepackage{amsmath}
\usepackage{mathtools}
\usepackage{txfonts}
\usepackage{mathdots}
\usepackage[normalem]{ulem}
\usepackage[dvips]{graphicx}
\usepackage{epsfig}
\usepackage{graphicx}
\usepackage{array}
\usepackage{amssymb}
\usepackage{amsfonts}
\usepackage{amsmath}
\usepackage{mathrsfs}
\usepackage{booktabs}
\usepackage{threeparttable}
\usepackage{multirow}
\usepackage{subfigure}
\usepackage{epsfig}
\usepackage{threeparttable}
\usepackage{chngpage}
\usepackage{float}
\usepackage{xcolor}
\usepackage{bm}
\usepackage{graphicx}
\usepackage[toc]{appendix}
\usepackage{tabularx}
\usepackage{graphicx}
\usepackage{adjustbox}
\usepackage{comment}
\usepackage{hyperref}
\usepackage{tabularx}
\usepackage{graphicx}
\usepackage{adjustbox}

\hypersetup{
    unicode=false,     
    pdftoolbar=false,  
    pdfmenubar=true,   
    pdffitwindow=false, 
    pdfstartview={FitH},
    pdftitle={},    
    pdfauthor={Authors},     
    pdfsubject={},   
    pdfcreator={},   
    pdfproducer={}, 
    pdfkeywords={quantum many-body scars} {superconducting processor} {quantum state tomography}, 
    pdfnewwindow=true,
    colorlinks=true,
    linkcolor=black,
    citecolor=blue, 
    filecolor=magenta,
    urlcolor=blue
}

\begin{document}

\title{Study on many-body phases in Jaynes-Cummings-Hubbard arrays}
\author{Jin-Lou Ma}
\thanks{These two authors contributed equally}
\author{Bobo Liu}
\thanks{These two authors contributed equally}
\affiliation{School of Physics, and Interdisciplinary Center for Quantum Information, Zhejiang University, Hangzhou $310027$, China}

\author{Qing Li}
\affiliation{School of Jia Yang, Institute for Quantum Technology and Engineering Computing, Zhejiang Shuren University, Shaoxing, Zhejiang $312028$, China}

\author{Zexian Guo}
\affiliation{School of Physics, and Interdisciplinary Center for Quantum Information, Zhejiang University, Hangzhou $310027$, China}

\author{Lei Tan}
\email{tanlei@lzu.edu.cn}
\affiliation{Lanzhou Center for Theoretical Physics, Key Laboratory of Theoretical Physics of Gansu Province, Lanzhou University, Lanzhou, Gansu $730000$, China}

\author{Lei Ying}
\email{leiying@zju.edu.cn}
\affiliation{School of Physics, and Interdisciplinary Center for Quantum Information, Zhejiang University, Hangzhou $310027$, China}

\begin{abstract}
Disorder in one-dimensional (1D) many-body systems emerges abundant phases such as many-body localization (MBL), and thermalization.  However, it remains unclear regarding their existence and behavior within hybrid quantum systems. Here, based on a simple bosonic-spin hybrid model, as known as the Jaynes-Cummings-Hubbard (JCH) array, we investigate the effect of disorder comparing to the phenomena in the clean system with the variation of atom-photon coupling strength. By using the level-spacing ratio, entanglement entropy, and the properties of observable diagonal and off-diagonal matrix elements, we find that strong disorder results in the appearance of MBL phase in the JCH model that strongly violate eigenstate thermalization hypothesis (ETH), while a conditional prethermal behavior can exist in weak disorder or weak coupling regime. The conditional prethermal dynamics is based on the choice of initial product states. This work systematically reveals abundant many-body phases in the 1D JCH model and clarifies the discrepancies in the thermalization properties of systems with and without disorder.
\end{abstract}

\maketitle
\section{INTRODUCTION}
Many strongly correlated particles in a closed quantum system enable abundant interesting phases. In past two decades, a large number of work conclude that existence of strong disorder can strongly violate ETH in 1D nonintegrable many-body systems and lead to a thermal-MBL phase transition in various systems~\cite{rigol2008thermalization,PhysRevLett.103.100403,PhysRevA.90.033606,
PhysRevA.80.053607,PhysRevB.99.155130,PhysRevE.93.032104,PhysRevE.87.012118,PhysRevE.89.042112,
PhysRevE.90.052105,PhysRevLett.120.200604,PhysRevLett.111.050403,PhysRevE.96.012157,PhysRevE.82.031130,
PhysRevLett.112.130403,PhysRevLett.122.070601,PhysRevLett.122.070601,PhysRevLett.124.040603}. The localized integrals of motion lead to the system retains information about its initial state for a long time in this dynamic phase. The significance of studying such systems lie in the order of excited states in its entire energy spectrum~\cite{Bauer_2013,PhysRevB.88.014206,Parameswaran_2018,PhysRevB.89.144201,
PhysRevLett.113.107204,PhysRevX.4.011052}, which were argued that it is potentially applied to the storage of quantum information~\cite{ALET2018498,doi:10.1146/annurev-conmatphys-031214-014726}.

The localization phase of many-body systems caused by disorder in a chain configuration for spins, fermions, or bosons has received extensive attentions~\cite{rigol2008thermalization,PhysRevLett.103.100403,PhysRevA.90.033606,
PhysRevA.80.053607,PhysRevB.99.155130,PhysRevE.93.032104,PhysRevE.87.012118,PhysRevE.89.042112,
PhysRevE.90.052105,PhysRevLett.120.200604,PhysRevLett.111.050403,PhysRevE.96.012157,PhysRevE.82.031130,
PhysRevLett.112.130403,PhysRevLett.122.070601,PhysRevLett.122.070601,PhysRevLett.124.040603}.A question asks whether such a MBL phase and its transition to thermalization can exist in a hybrid quantum many-body systems, in which the JCH model is a typical example that has the advantages of precise manipulation, individual addressing, and the construction of any geometric structure~\cite{https://doi.org/10.1002/lpor.200810046}. The JCH model is a hybrid system of photons and spins, in which there are novel matter states and  phenomena~\cite{https://doi.org/10.1002/lpor.200810046}.

\begin{figure}[tbh]
\centering
\includegraphics[width=8.5cm]{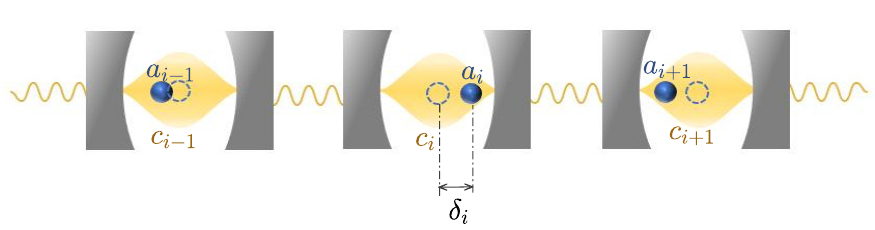}
\caption{
Schematic diagram of the JCH array with coupling disorder. The 1D JCH model comprises interconnected Jaynes-Cummings models via photon tunneling (represented by yellow wavy curves), wherein each  cavity accommodates a two-level atom (depicted by blue ball), which is randomly fixed at a position $\delta_i$ deviating from the central position of the cavity.
}\label{fig1}
\end{figure}

In this paper, we focus on the JCH model formed by a set of coupled cavities to trap photons interacting with two-level atoms or qubits. Such a system is non-integrable~\cite{Li_2021,Li_2021b}. For experimental platforms, disorder is inevitable. It is not yet clear how disorder affects the thermalization properties of the JCH model.
Here, we choose the atom-photon interaction as the disordered quantity, induced by the random locations of the atom in cavities, as illustrated in Fig.~\ref{fig1}. Thus, the disorder does not change the sign of the atom-cavity coupling strength, being in a range of $[0,D]$, where $D$ is the maximum coupling strength. In this paper, we will unravel the veil of the influence of this kind of disorder on the thermal properties of the JCH chain and discuss its difference from a clean system.

Through the numerical simulation, we find that under the weak disorder strength related to atom-photon interactions, the system behaves in the quasi-MBL phase as if there was no disorder~\cite{PhysRevLett.117.240601}, and shows a prethermal period dependent on the initial state. At the regime of medium disorder strength, ergodic phase appears in both disordered and clean systems. At strong coupling strength, accompanied by strong disorder, the system enters into a typical MBL phase, while in the clean system, there is an emergence of quasi-MBL phase. The paper is organized as below. Sec. ~\ref{II} presents the theoretical model and introduces level-
spacing ratio and entanglement entropy to study the various phases of the disordered JCH system. We investigate the effect of disorder on phase transition in Sec. ~\ref{III}. Sec. ~\ref{IV} is devoted to discussion of the eigenstate thermalization of the disordered and the clean JCH systems.

\section{Disordered and clean Jaynes-Cummings-Hubbard arrays}\label{II}

We consider a disordered 1D JCH model and the schematic diagram of it is shown in Fig. ~\ref{fig1}~\cite{Mascarenhas_2012}, whose Hamiltonian at the rotating wave approximation is given by ($\hbar=1$)
\begin{align}
\begin{split}
H^{\prime}=& \sum^L_{i}\left[\omega_ca_i^{\dagger}a_i+\omega_a\sigma^{+}_{i}\sigma^{-}_{i}+g_i\left(a_i\sigma^{+}_{i}+a_i^{\dagger}\sigma^{-}_{i}\right)\right]{}\\ &{}-J\sum^{L-1}_i \left(a_i^{\dagger}a_{i+1}+a_ia^{\dagger}_{i+1}\right),\label{H01}
\end{split}
\end{align}
where the first term of the Hamiltonian describes free Hamiltonians of photons and two-level atom system on each site, $\omega_a$ ($\omega_c$) is the frequency of the two-level atom (photons) in single cavities. We only consider the resonance frequency case ($\omega_a=\omega_c$). $\sigma_i^{-}$ and $\sigma_i^{+}$ are the atomic raising and lowering operators, respectively. $a_i^{\dagger}$ ($a_i$) is the photon creation (annihilation)
operator for the $i$th site. $L$ is the number of lattice sites. The atom-photon coupling strength $g_i\in[0,D]$ for $i$th cavity and $D$ denotes the disorder strength~\cite{PhysRevA.101.053805}.
The second term is the sum of a hopping term of photons and we assume that all the hopping strength of photons between the nearest neighboring cavities is identical and equal to $J$.
By using the rotating transformation operator $U=\mathrm{exp}\left[-i\sum^L_{j=1}\omega_c( a_j^{\dagger}a_j+\sigma^{+}_{j}\sigma^{-}_{j})t\right]$, the Hamiltonian in Eq.~(~\ref{H01}) can be re-written as
\begin{align}
\begin{split}
H=& \sum^L_{i}g_i\left(a_i\sigma^{+}_{i}+a_i^{\dagger}\sigma^{-}_{i}\right)-J\sum^{L-1}_i \left(a_i^{\dagger}a_{i+1}+a_ia^{\dagger}_{i+1}\right).\label{H02}
\end{split}
\end{align}
Numerical simulation in the rest of content is based on Hamiltonian Eq.~(~\ref{H02}).
In addition, the total number of atomic and photonic excitations is fixed as $N=\sum_{i}(a_i^{\dagger}a_i+\sigma^{+}_{i}\sigma^{-}_{i})=\sum_{i}(n_i^{\mathrm{c}}+n_i^{\mathrm{a}})$, and we consider the open boundary condition (OBC) with the filling factor is $\nu \equiv N/L=1/2$. The disordered JCH model only has a chiral symmetry and its corresponding chiral operator is~\cite{Li_2021b} $\Gamma=\Pi_{j\in \mathrm{even}}e^{i\pi a^{\dag}_ja_j}\Pi_{j\in \mathrm{odd}}\sigma_j^{z}$.
Then, the dimension of Hilbert space of $H$ is given by~\cite{Li_2021b,PhysRevB.105.165432}
\begin{align}
\mathcal{D}=\sum_{s=1}^{N}\frac{L(N+L-s-1)!}{(N-s)!(L-s)!s!}.
\end{align}
 The basis vectors are written as $|\mathbf{n}\rangle\equiv \prod_{i}|n_i^{\mathrm{c}},n_i^{\mathrm{a}}\rangle_i$. By utilizing the exact diagonalization, the maximum size of the system is $L = 10$.

In order to further clarify the special behaviors of disorder, the results of the disordered JCH model are compared to those of clean JCH model. The Hamiltonian of the clean JCH model is given by
\begin{equation}
\begin{split}
H_{\mathrm{cl}}=g_{\mathrm{cl}}\sum^L_{i}(a_i\sigma^{+}_{i}+a_i^{\dagger}\sigma^{-}_{i})-J\sum^{L-1}_i (a_i^{\dagger}a_{i+1}+a_ia^{\dagger}_{i+1}).
\end{split}\label{eq:H_cl}
\end{equation}
The Hamiltonian $H_{\mathrm{cl}}$ owns the  extra  reflective symmetry. Under the reflection (parity) operator $P$, we study the clean JCH model in antisymmetric subspaces.

For characterized MBL phase and ergodic phase, we need to introduce two physical quantities. The first one is the statistical features of spectrum by the level-spacing ratio $\langle r \rangle$~\cite{PhysRevB.75.155111}, which is a statistical quantity and is the average over $r_n=\mathrm{min}\left\{\Delta E_{n+1}/\Delta E_n,\Delta E_n/\Delta E_{n+1}\right\}$ with $\Delta E_n=E_{n+1}-E_n$. Here, $E_n$ is the $n$th eigenenergy, chosen by the middle third of the energy spectrum. For the MBL phase, the level-spacing ratio exhibits a Poisson distribution with $\langle r\rangle\approx0.386$, while it shows the Wigner-Dyson distribution with $\langle r\rangle\approx0.536$ in the ergodic phase. In this paper, `` $\langle \cdot\rangle$ '' indicates the average of physical quantities including eigenstates and disordered realizations. The other quantity is the half-chain entanglement entropy (EE) $S_{L/2}=-\mathrm{Tr}\left[\rho_s\mathrm{log}(\rho_s)\right]$ with $\rho_s=\mathrm{Tr}_{i\leq{L}/{2}}\left[|n\rangle\langle n|\right]$. The EE describes how information spreads from one part of the system~\cite{doi:10.1146/annurev-conmatphys-031214-014726}.
In the MBL phase, the average EE, $\langle S_{L/2}\rangle$, slowly grows as the time evolution and follows an area-law scaling~\cite{PhysRevLett.113.107204,Bauer_2013,PRXQuantum.3.030201}. Differently, $\langle S_{L/2}\rangle$ yields a volume-law scaling in the ergodic phase, which approaches to the Page value $S_{\mathrm{P}}$ for a random pure state~\cite{PhysRevLett.71.1291}. To clearly describe the occurrence of MBL to ergodic phase transition, it is also necessary to show the sample-to-sample deviation of the half-chain EE $\Delta{S}$,
and its peak value represents the phase transition point~\cite{PhysRevLett.113.107204,PhysRevLett.119.075702,PhysRevX.7.021013}.

\section{Many-body phases and transitions}\label{III}

\begin{figure}[tbh]
\centering
\includegraphics[scale=0.45,trim=0 0 0 0]{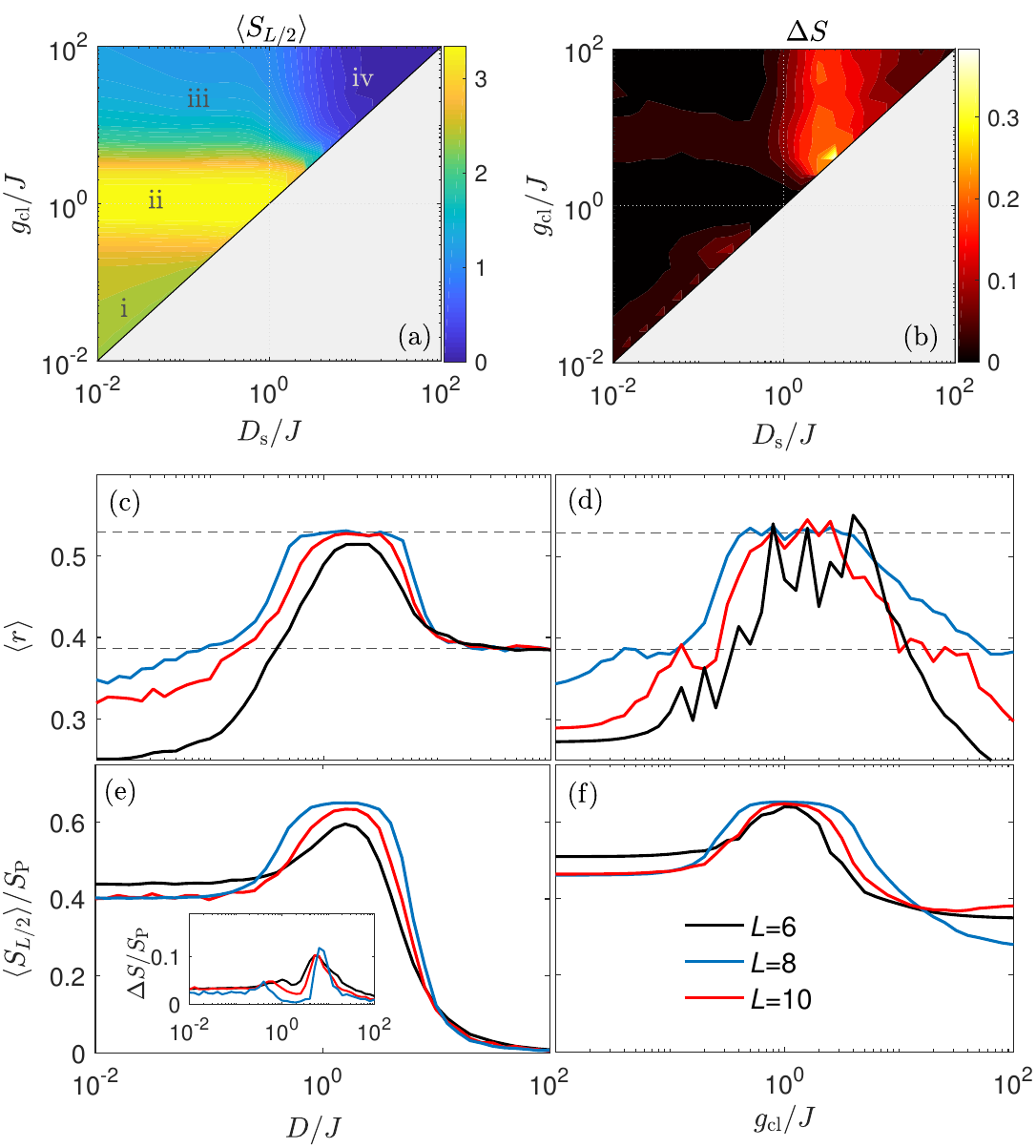}
\caption{ (a) The EE $\langle S_{L/2}\rangle$ and (b) the sample-to-sample deviation of the EE $\Delta{S}$ as a function of $g_{\mathrm{cl}}/J$ and $D_\mathrm{s}/J=g_i/J-D/2J$, respectively. The level-spacing ratio $\langle r\rangle$ (c) and the EE $\langle S_{L/2}\rangle /S_{\mathrm{P}}$ (e) as functions of the scaled disorder strength $D/J$ in the disordered JCH model. The insets of Fig. ~\ref{fig2} (e) is the deviation of the EE $\Delta{S}/S_{\mathrm{P}}$. The average half-chain EE is rescaled by the page value $S_{\mathrm{P}}$. (d) show the level-spacing ratio $\langle r\rangle$ as functions of the scaled coupling strength $g_{\mathrm{cl}}/J$ in the clean JCH model. The  EE $\langle S_{L/2}\rangle /S_{\mathrm{P}}$ as functions of the scaled coupling strength $g_{\mathrm{cl}}/J$ in Fig. ~\ref{fig2} (f). Grey dashed lines mark $\langle r\rangle =0.386$ (Poisson distribution) and $0.536$ (Wigner-Dyson distribution) in Fig. ~\ref{fig2} (c) and (d). In the clean JCH model, the photon-atom coupling strength $g_{\mathrm{cl}}$ of each onsite is assumed to be the same. The averaged physical quantities of the disordered JCH model are taken by $1000$, $400$, $50$ disordered samples for $L=6$, $8$, $10$.}\label{fig2}
\end{figure}

\begin{figure}[tbh]
\centering
\includegraphics[height=6.8cm]{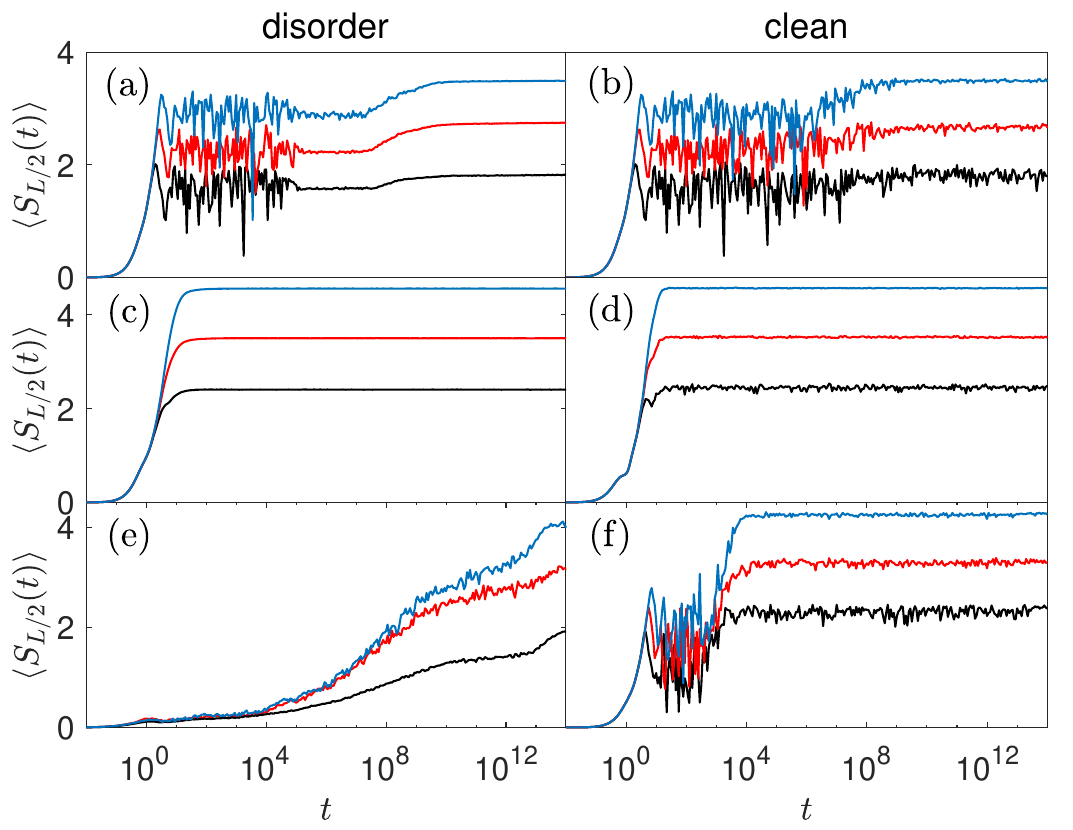}
\caption{ The average half-chain EE $\langle S_{L/2}(t)\rangle$ versus time $t$ for the disordered (right) and the clean (left) JCH model. Three colors represent three sizes $L=6$ (black), $8$ (red), $10$ (blue). The top, middle and bottom figures correspond to $D/J,g_{\mathrm{cl}}/J=0.01$, $2$ and $100$, respectively. The averaged physical quantities of the disordered JCH model are taken by $1000$, $400$, $100$ disordered samples for $L=6$, $8$, $10$. Note that, we plot the average time evolution of $3\langle S_{L/2}(t)\rangle$ in Fig.~\ref{fig3}(e).}\label{fig3}
\end{figure}

\begin{figure}[h]
\centering
\includegraphics[scale=0.5,trim=50 210 0 250]{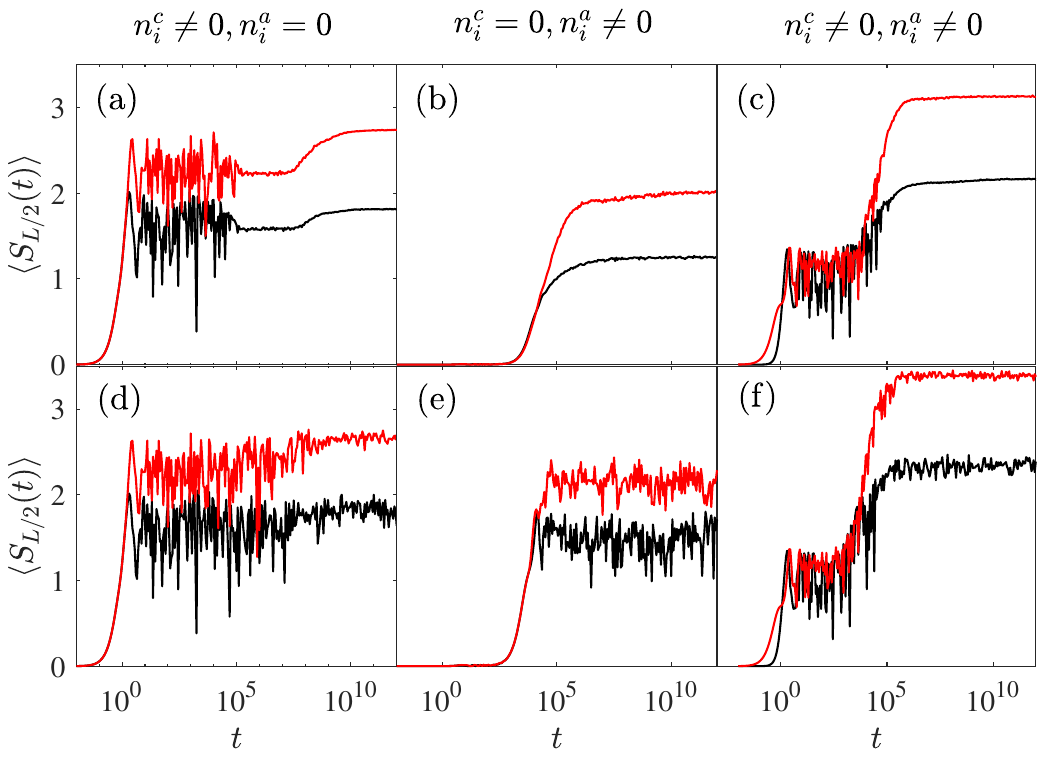}
\caption{ The average half-chain EE $\langle S_{L/2}(t)\rangle$ versus time $t$ for the disordered (First column) and the clean (Second colum) JCH model with different initial states. The disordered and clean cases correspond to $D/J$, $g_{\mathrm{cl}}/J = 0.01$, respectively. Fig.~\ref{fig4} (a) and (d) are shown by the initial state ($N^{\mathrm{c}}\neq0$, $N^{\mathrm{a}}=0$). Fig.~\ref{fig4} (b) and (e) are shown by the initial state ($N^c\neq0$, $N^a\neq0$). Fig.~\ref{fig4} (c) and (f) are shown by the initial state ($N^{\mathrm{c}}=0$, $N^{\mathrm{a}}\neq0$). $N^{\mathrm{c}}$ ($N^{\mathrm{a}}$) is the number of photonic (atomic) excitations. The black and red lines represent sizes $L=6$, $8$, respectively. The $1000$ ($100$) disordered samples correspond to $L=6$ ($8$) for the disordered case.}\label{fig4}
\end{figure}

\begin{figure*}
\centering
\includegraphics[width=\linewidth]{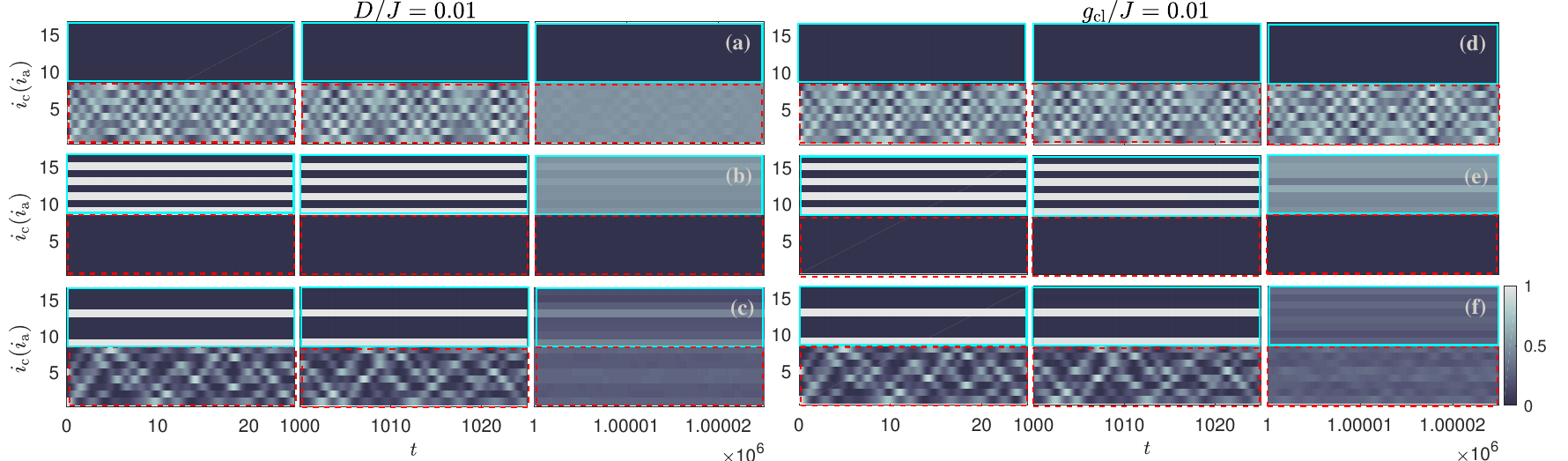}
\caption{ The averaged atomic (photonic) excitations occupation occupancy per site $\langle n^a_i \rangle$ ($\langle n^{\mathrm{a}}_i \rangle$) as a function of time $t$ with different initial states for the disordered and clean JCH models. The initial state of Fig.~\ref{fig5} (a) and (d) is $\prod_{i\in \mathrm{odd}}|1,g\rangle_i\otimes\prod_{j\in \mathrm{even}}|0,g\rangle_j$, the initial state of Fig.~\ref{fig5} (b) and (e) is $|1,e\rangle_1\otimes|1,e\rangle_5\otimes\prod_{j\in \mathrm{others}}|0,g\rangle_j$ and the initial state of Fig.~\ref{fig5} (c) and (f) is $\prod_{i\in \mathrm{odd}}|0,e\rangle_i\otimes\prod_{j\in \mathrm{even}}|0,g\rangle_j$. The site index of photonic (atomic) excitations is $1$ to $8$ ($9$ to $16$). Each initial state exhibits three periods of time evolution, namely $0\leq t_1<25$, $1000\leq t_2<1025$ and $10^6\leq t_3<1.000025*10^6$. The disordered and clean cases correspond to $D/J$, $g_{\mathrm{cl}}/J = 0.01$, respectively. The system size is chosen by $L=8$. The averaged physical quantities of the disordered JCH model are taken by $100$ disordered samples.}\label{fig5}
\end{figure*}

We set out to the responses of the level-spacing ratio $\langle r\rangle$ and the average half-chain EE of disordered (clean) model with increasing the disorder strength $D$ (the pristine photon-atom coupling strength $g_{\mathrm{cl}}$). Figures ~\ref{fig2} (a) and (b) depict the average EE, $\langle S_{L/2}\rangle$, and its deviation, denoted as $\Delta{S}$, respectively, as a function of the coupling strength $g_{\mathrm{cl}}$ and disorder strength $D_\mathrm{s}/J=g_i/J-D/2J$. There are three many-body phases in the disordered JCH model, i.e. quasi-MBL, MBL, and ergodic phases.
In Figs. ~\ref{fig2} (c)-(f), as the disorder strength increases, the level-spacing ratio exhibits a range of distributions, transitioning from the quasi-Poisson distribution to the Wigner-Dyson distribution, and finally converging to the Poisson distribution.
Similarly, the EE  undergoes a transition from a quasi-volume-law, then a volume-law behavior, to an area-law behavior eventually.
As for the clean JCH model, these two quantities exhibit similarities to those of the disordered JCH model in the regime of weak and intermediate disorder strengths. However, for strong disorder strengths, the finite-size effect is almost negligible, whereas for strong coupling interactions without disorder, the finite-size effect is relatively pronounced. We also show the sample-to-sample deviation of the  EE $\Delta{S}/S_{\mathrm{P}}$ for the disordered JCH model in the inset of Fig.~\ref{fig2} (e).
The enhancement of the peak value of $\Delta{S}/S_{\mathrm{P}}$ at $D/J\sim 10^1$ is at larger system size $L=10$, implying that the system shows a ergodic-MBL phase.
Note that the value of the other peak ($D/J\sim 10^0$)  approaches to the weak disorder strength with the increase of size $L$. Thus, we suppose that, under the weak disorder limit, the disordered system presents the same integrable behaviors as for clean system~\cite{PhysRevB.105.165432}. Based on above results, it can be concluded that the intermediate disorder displays an ergodic phase, while the strong disorder presents a MBL phase.

Next, we show the dynamics of the average half-chain EE for different many-body phases in the disordered and the clean JCH model. Previous work indicate that the EE dynamics shows a  scaling behavior of $\mathrm{log} t$ for the MBL phase~\cite{PhysRevB.77.064426,PhysRevLett.109.017202}, while the EE rapidly tends to a saturation value in the ergodic phase~\cite{PhysRevB.102.094201}. Here, Fig.~\ref{fig3} shows the time evolution of the EE under different parameters, where the initial state is chosen as $|\mathbf{n}\rangle_{\mathrm{in}}\equiv \prod_{i\in \mathrm{odd}}|1,g\rangle_i\otimes\prod_{j\in \mathrm{even}}|0,g\rangle_j$. In Fig.~\ref{fig3} (a), we can find that at a disorder strength of $D/J=0.01$, its EE exhibits a rapid increase at early time, followed by oscillations, a metastable period, and eventually approaches a saturation value slowly. One can find that the time-average values of  $\langle S_{L/2}(t)\rangle$ for oscillating and metastable period regimes are almost identical. This phenomenon is similar to that of the weak disordered spin ladder system~\cite{PhysRevB.102.094201}. It can be seen from Fig.~\ref{fig3} (a) that the phenomena of oscillation and metastable period remain stable across different system sizes. As the number of disordered samples increase, the oscillating period tends to become invariant and the metastable period becomes a smooth function (see details in Fig.~\ref{fig9} (a) of the APPENDIX A). This observation suggests that oscillation is a inherent characteristic of the weak disorder system, while the average behavior of disordered realizations gives rise to a metastable period in the finite-size system. In addition, at disorder-free case with a small coupling $g_{\mathrm{cl}}/J=0.01$, the EE dynamics show similar phenomena to the weak disorder case, while the phenomenon of metastable period disappears, as shown in Fig. ~\ref{fig3} (b). The observed difference can be interpreted as an indication that weak disorder or coupling strength induces prethermalization during the oscillating and metastable periods~\cite{Mori_2018,PhysRevLett.115.180601,PhysRevB.102.094201}.

When $D/J$, $g_{\mathrm{cl}}/J=2$, both disordered and clean cases are in ergodic phases, the EE tends to reach saturation rapidly. In the disordered system at $D/J=100$ (Fig. ~\ref{fig3} (e)), the EE $\langle S_{L/2}(t)\rangle$ exhibits a $\mathrm{log}(t)$ scaling behavior before reaching a saturation value. While for the clean system at $g_{\mathrm{cl}}/J=100$ [Fig. ~\ref{fig3} (f)], the results are similar to the case of $g_{\mathrm{cl}}/J=0.01$. The difference lies in the fact that the latter one has a  prolonged oscillating prethermalization for the certain initial states and ultimately reaches a saturation value rapidly within the finite-size system. It is noteworthy that the dynamics of the EE differ significantly between the disordered and clean coupling interactions under strong disorder conditions. For the moment, we roughly consider that the weak $D/J$, $g_{\mathrm{cl}}/J$ and the strong $g_{\mathrm{cl}}$ are quasi-MBL phases, the intermediate regime of $D/J$ and $g_{\mathrm{cl}}/J$ are ergodic phases and the strong $D/J$ is a MBL phase.

Furthermore, We find that the emergence of prethermal dynamics at weak disorder $D/J=0.01$ (weak coupling $g_\mathrm{cl}/J=0.01$) regime strongly depends on  initial states. Figure~\ref{fig4} shows that the dynamics of half-chain EE $\langle S_{L/2}(t)\rangle$ for different initial states with the weak disorder strength $D/J$ (coupling strength $g_{\mathrm{cl}}/J$) being equal to $0.01$. The prethermal dynamics occurs, for the initial state with only photonic excitations, while the EE would rapidly grows after a long threshold time for the initial states with only atomic excitations. If the initial state is consisting of both atomic and photonic excitations, the dynamics of half-chain EE also shows a prethermal behavior. Differently, its prethermal regime is shorter and the thermal plateau is higher than the case of initial states with only atomic excitations.

To reveal the reason of these differences, we also plot the disorder-average population dynamics for different kinds of initial states at weak disorder $D/J=0.01$ regime in Fig.~\ref{fig5}. The whole JCH system is consisting of atomic and photonic parts. For the initial state only with photonic excitations, the populations are constrained in the photonic part, exhibiting a prethermal dynamics, distinguishing from the case of clean system ($g_\mathrm{cl}/J=0.01$) with a non-thermal dynamics, as shown in Figs.~\ref{fig5} (a) and (d). For the initial states only with atomic excitations, the populations stay at the atomic part, showing a localized dynamics in both cases of disordered and clean systems (see in Figs.~\ref{fig5} (b) and (e)). If the initial states with both atomic and photonic excitations,  both atomic and photonic parts exhibit prethermal dynamics, implying the larger entanglement entropy than the case of the initial states only with photonic excitations.


\section{Eigenstate thermalization properties}\label{IV}

\begin{figure*}[t]
\centering
\includegraphics[width=\linewidth]{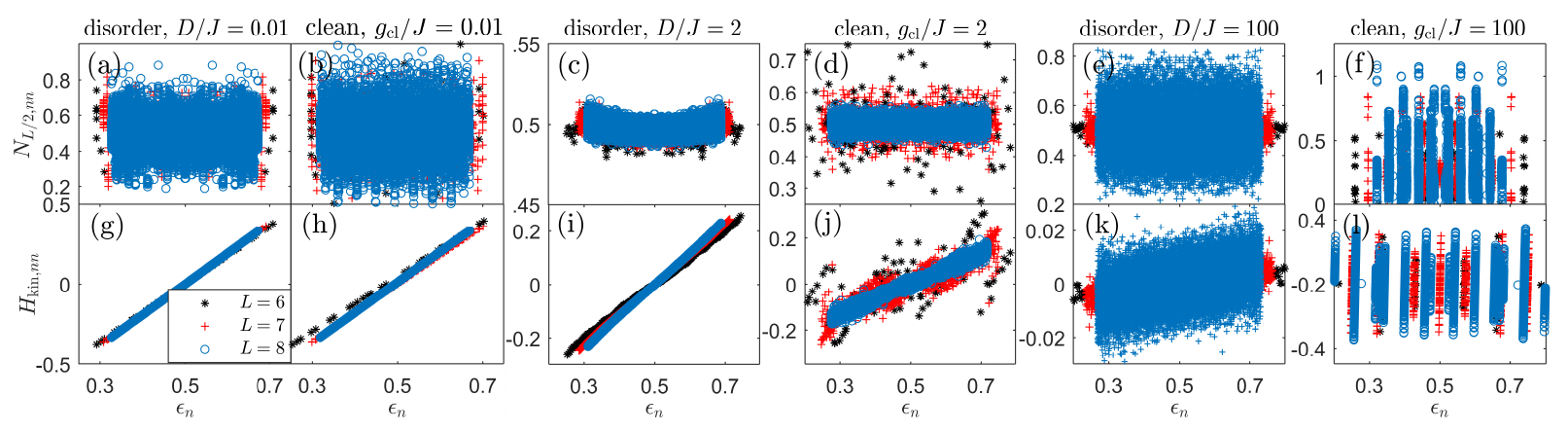}
\caption{ The diagonal matrix elements of $N_{L/2}$ (a)-(f) and $H_{\mathrm{kin}}$ (g)-(l) as a function of the energy density $\epsilon_n$ with different $D/J$ and $g_{\mathrm{cl}}/J$ for the disordered and clean JCH models. The black-star, red-plus, blue-circle lines correspond to $L=6$, $8$, $10$, respectively. The averaged physical quantities of the disordered JCH model are taken by $1000$, $400$, $100$ disordered samples for $L=6$, $8$, $10$, respectively.}\label{fig6}
\end{figure*}

To further investigate the entire system properties, in this section, we analyze the thermalization of the disordered and clean JCH model to examine the validity of the ETH in different many-body phases which were studied by the EE dynamics above. In order to determine whether the system can be thermalized, it is necessary to pay attention to whether the diagonal and non-diagonal elements of the local observable operator $O$ satisfies the ETH~\cite{d2016quantum,Srednicki_1999,PhysRevA.43.2046,PhysRevE.50.888}. The definition of local operator is written as
\begin{equation}\label{H04}
O_{nm}=O(\bar{E})\Delta E_{nm}+e^{-S(\bar{E})/2}f_{O}(\bar{E},\omega^{\prime})R_{nm},
\end{equation}
where $\bar{E}=(E_{n}+E_{m})/2$ is the average energy of adjacent eigen energies, $\omega^{\prime}=E_{n}-E_{m}$ is its energy difference. Here, $n$ and $m$ are the indices of  eigen states, $S(\bar{E})_\mathrm{th}$ stands for the thermodynamic entropy, and $R_{nm}$ is the random variable which obeys a normal distribution. In the thermodynamic limit, $O(\bar{E})$ and $f(\omega^{\prime},\bar{E})$ are smooth functions about $\omega^{\prime}$ and $\bar{E}$. The first term in Eq.~(\ref{H04}) is the expected values of the observable and the second term describes the off-diagonal matrix elements. The average eigenstate-to-eigenstate fluctuations of diagonal expectation is given by~\cite{PhysRevE.90.052105}
\begin{align}
\overline{|\delta O_{nn}|}= \overline{|O_{n+1,n+1}|-|O_{n,n}|}.\label{H05}
\end{align}
In general, the eigenstate-to-eigenstate fluctuations both for diagonal and off-diagonal elements exponential decay as the system size increases if the system satisfies the ETH~\cite{PhysRevE.87.012118,PhysRevE.90.052105,PhysRevE.89.042112,PhysRevE.93.032104,
PhysRevLett.120.200604,PhysRevB.99.155130,PhysRevE.100.062134}.
Here, we select two specific local observables to discuss whether their behaviors are consistent with the predictions of the ETH. The first observable is the occupancy operator $N_{L/2}$ at the site $L/2$, while the second observable is the kinetic operator per site $H_{\mathrm{kin}}=({1}/{L})\sum^{L-1}_i(a_i^{\dagger}a_{i+1}+a_ia^{\dagger}_{i+1})$, which represents the reduced photon hopping term.

\begin{figure}[h]
\centering
\includegraphics[width=\columnwidth]{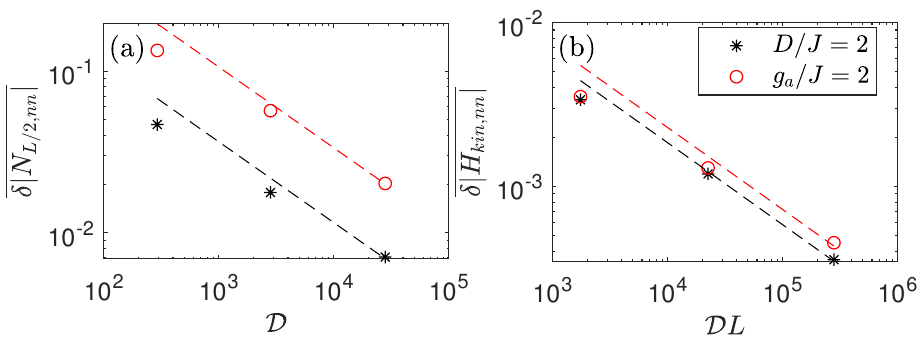}
\caption{ Scaling of (a) $\overline{|\delta N_{L/2,nn}|}$  and (b) $\overline{|\delta H_{\mathrm{kin},nn}|}$ at the nonintegrable point of the disordered ($D/J=2$) and the clean ($g_{\mathrm{cl}}/J=2$) JCH model. The dashed lines denote a power law scaling of $\propto x^{-1/2}$ in Figs. ~\ref{fig7} (a) and (b). The average physical quantities of the disordered JCH model are taken by $1000$, $400$, $100$ disordered samples for $L=6$, $8$, $10$, respectively. }\label{fig7}
\end{figure}

Firstly, the diagonal elements of observable $N_{L/2}$ and $H_{\mathrm{kin}}$ as functions of the energy density are plotted in Fig. ~\ref{fig6}. The energy density defined by $\epsilon_n=(E_n-E_\mathrm{min})/(E_\mathrm{max}-E_\mathrm{min})$, where $E_n$ is the $n$th eigenenergy, $E_\mathrm{min}$ ( $E_\mathrm{max}$) represents the minimum (maximum) eigenenergies. Here, we focus on the middle four-fifths of the energy spectrum.
In Figs. ~\ref{fig6} (a,b) and (g,h), it can be seen that, at disorder strength $D/J = 0.01$ and coupling strength $g_{\mathrm{cl}}/J = 0.01$, the fluctuations of the disordered and clean cases do not diminish with increasing system size $L$ both for the observables $N_{L/2}$ and $H_{\mathrm{kin}}$. For the observable $H_{\mathrm{kin}}$, whether it is the disordered or clean case, we find that with the increase of energy density $\epsilon_n$, the expected value changes linearly with minor fluctuation. The result indicates that the atom-photon coupling term acts as a small perturbation, the hopping term of the photon and the Hamiltonian $H$ of Eq.~\ref{H02} can be regarded as commutative. This implies that the observable $H_{\mathrm{kin}}$ and the Hamiltonian in Eq. ~\ref{H02} share almost identical eigenvalues and eigenstates. Therefore, $H_{\mathrm{kin}}$ is a linear function of the energy density $\epsilon_n$. Thus, $H_{\mathrm{kin}}$ cannot be simply considered as a local observable to diagnose the thermalization. At  a mediate disorder strength, say $D/J= 2$, the fluctuations of the observable $N_{L/2}$ decrease as the size $L$ enlarges. But, the observable $H_{\mathrm{kin}}$ is almost a smooth function of energy density $\epsilon_n$, even in small system sizes. The behaviors of the clean system ($g_{\mathrm{cl}}/J= 2$) is consistent with that of the disordered system. Also, we can  see this phenomenon from the average eigenstate-to-eigenstate fluctuations $\overline{|\delta O_{nn}|}$ of diagonal elements decreases exponentially fast with increasing $L$ in Fig. ~\ref{fig7} (a) and (b) for ergodic phases in the disordered and clean systems. Due to the Hilbert-Schmidt norm of  operator $H_{\mathrm{kin}}$ scales as $1/\sqrt{L}$~\cite{PhysRevE.100.062134,PhysRevLett.124.040603}, the average eigenstate-to-eigenstate fluctuations of $H_{\mathrm{kin}}$ to be $\propto(L\mathcal{D})^{-1/2}$. In the case of $D/J$, $g_{\mathrm{cl}}/J=100$, although the fluctuations of the two observables increase with the increasing of system size for two kinds of systems, it is remarkable that the diagonal elements of disordered and clean systems change differently with energy densities. Specifically, the expected values of the observables show a uniform distribution for disordered systems. while for clean system, there is a large amount of quasi-degeneracy in the energy densities,  resembling the separation of energy bands. In short, by comparing the distributions of diagonal elements between disordered and clean systems, we find that the fluctuations in the disordered case are noticeably smaller, in particular in the ergodic phases ($D/J,g_{\mathrm{cl}}/J=2$). This discrepancy can be attributed to the averaging effect of the disordered samples.



Based on Fig.~\ref{fig6} and Fig.~\ref{fig7}, we can conclude that  $D/J$, $g_{\mathrm{cl}}/J=2$ (ergodic phase) meets ETH, while $D/J=0.01$, $g_{\mathrm{cl}}/J = 0.01$, $100$ (quasi-MBL phase) and $D/J= 100$ (MBL phase) strongly violate ETH. We also find that the average disordered diagonal elements of $N_{L/2}$ and $H_{\mathrm{kin}}$  are symmetrical about the axis of $\epsilon_n= 0.5$ and the point $(\epsilon_n,H_{\mathrm{kin},nn})=(0.5,0)$, respectively. This symmetry arises due to the commutation relation $[\Gamma,N_{L/2}]=0$, which leads to $\langle n | N_{L/2}| n \rangle=\langle n |\Gamma^{\dag} N_{L/2}\Gamma| n \rangle$. In addition, the anticommutation relation $\{\Gamma,H_{\mathrm{kin}}\}=0$ results in  $\langle n | H_{\mathrm{kin}}| n \rangle=-\langle n |\Gamma^{\dag} H_{\mathrm{kin}}\Gamma| n \rangle$~\cite{Li_2021b}. Differently, the diagonal elements of the clean JCH model are not symmetrical since the excitation number $N$ is odd. According to the symmetry analysis in the Appendix B, when the chiral operator $\Gamma$ and the reflection operator $P$ commute, the system has the chiral symmetry in the antisymmetric subspace with reflective symmetry for the even excitation number $N$. However, when the operators $\Gamma$ and $P$ do not commute, there is no such a chiral symmetry in the antisymmetric subspace for the odd excitation number $N$. For the disordered system, the reflection symmetry is lost, thus the chiral symmetry emerges in the systems both with odd and even excitations.

\begin{figure*}[t]
\centering
\includegraphics[scale=0.63,trim=0 0 0 0]{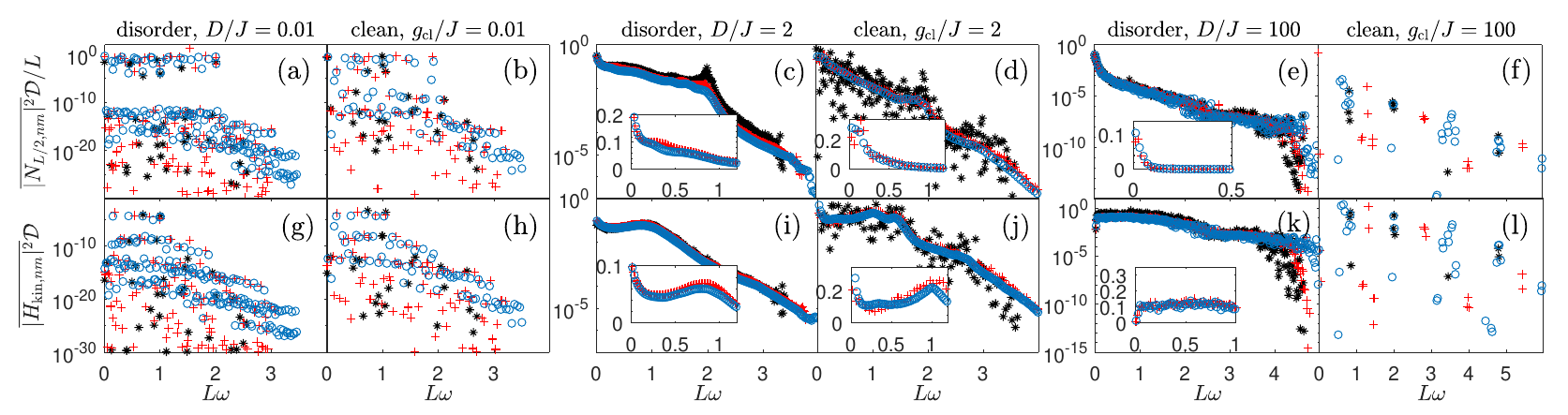}
\caption{ (a)-(f) Coarse-grained averages of $N_{L/2}$ as a function of $L\omega$ with different $D/J$ and $g_{\mathrm{cl}}/J$ for the disordered and clean JCH models. (g)-(l) Coarse-grained averages of $H_{\mathrm{kin}}$ as a function of $L\omega$ with different $D/J$ and $g_{\mathrm{cl}}/J$ for the disordered and clean JCH models.  The matrix elements are computed within a small window of energy around the average spectrum $\bar{\epsilon}$ of width $0.01\omega$. The averages in $\omega$ are calculated in windows with $\delta\omega=0.002$. The black-star, red-plus, blue-circle lines correspond to $L=6$, $8$, $10$, respectively. The averaged physical quantities of the disordered JCH model are taken by $1000$, $400$, $100$ disordered samples for $L=6$, $8$, $10$, respectively.}\label{fig8}
\end{figure*}

Here, we focus on the variance of the off-diagonal elements. In our model, the variance $
\left|\overline{O_{nm}}^2-\overline{|O_{nm}|^2}\right|\approx\overline{|O_{nm}|^2}$ since observables $\overline{H_{\mathrm{kin},nm}}\approx0$ and $\overline{N_{L/2,nm}}\approx0$, as same in spin systems~\cite{PhysRevE.100.062134,PhysRevLett.125.070605,PhysRevB.102.075127}. Also, $\overline{|O_{nm}|^2}$ is a quantity to study fluctuation dissipation relation~\cite{PhysRevLett.111.050403}, transport properties~\cite{PhysRevLett.117.170404,PhysRevE.87.012118}, periodic driven heating rate~\cite{PhysRevLett.123.240603}, etc. In Fig.~\ref{fig8}, we plot the coarse-grained average scaled variances $\overline{|H_{\mathrm{kin},nm}|^2}$ and $\overline{|N_{L/2,nm}|^2}$ of the off-diagonal matrix elements with $\omega=\epsilon_{n}-\epsilon_{m}$. For $D/J$, $g_{\mathrm{cl}}/J=0.01$, the properties of the two systems are similar, both of them have a strong dispersion. At disorder strength $D/J=2$ and coupling strength $g_{\mathrm{cl}}/J=2$, the coarse-grained averages $\overline{|H_{\mathrm{kin},nm}|^2}$ and $\overline{|N_{L/2,nm}|^2}$  of the off-diagonal matrix elements show smoothing functions of $\omega$. The variance of off-diagonal matrix elements satisfies $\overline{|O_{nm}|^2}\propto(L\mathcal{D})^{-1}$~\cite{PhysRevE.100.062134,PhysRevLett.124.040603}. The difference in scaling behaviors of the two observables can be attributed to the Hilbert-Schmidt norms of the observable $H_{\mathrm{kin}}$, whose scaling behaviors are given by $\sim 1/\sqrt{L}$. On the other hand, the off-diagonal matrix elements of the observables have similar behaviors in the disordered and clean systems. At strong disorder and strength coupling regimes, say $D/J$, $g_{\mathrm{cl}}/J=100$, the variances of two observables are the smooth functions of $\omega$ for the disordered systems, but not for clean systems. In the clean system, the behavior is similar to the case at weak coupling strength ($g_{\mathrm{cl}}/J$). In Fig. ~\ref{fig8}, we have that the variances of the observables for $L=8$ and $L=10$ show minimal finite-size effects in both the ergodic and MBL phases.

The scaled variances of the off-diagonal matrix elements in the low frequency $\omega$ part is briefly discussed below for the ergodic and MBL phases. Observables $N_{L/2}$ and $H_{\mathrm{kin}}$ exhibit data collapse as $L\omega$ decrease for different system sizes (See the insets of Fig. ~\ref{fig8}). For the the ergodic phases ($D/J$, $g_{\mathrm{cl}}/J=2$), the collapse degrades as $L\omega$ increases and two variances of observables have a high value as $L\omega$ approaches to zero, indicating the diffusive dynamics, as same with quantum-chaotic systems~\cite{d2016quantum}. In addition, for the MBL phases ($D/J=100$) with a large size, the variance of observable $N_{L/2}$ does not vanish as $L\omega$ approaches zero, while the observable $H_{\mathrm{kin}}$ approaches to be zero. This phenomenon is similar with the integrable
XXZ chain~\cite{PhysRevB.102.075127}. By comparing Fig. ~\ref{fig8}(c), ~\ref{fig8}(d) and ~\ref{fig8}(e), we find that the variance of observable $N_{L/2}$ has the same behavior in the low frequency regime for both the ergodic and MBL phases.  This implies that the scaling behavior of observable $N_{L/2}$ is stable in ergodic and MBL phases. However, as the relationship between the variances of the observables and the frequency $\omega$ is not a smooth function in other cases, we will not   discuss in depth.

To study the normality of distribution of the off-diagonal matrix elements, we calculate the ratio~\cite{PhysRevE.100.062134}
\begin{equation}\label{H06}
\Gamma_{O}(\omega)=\overline{|O_{n,m}|^2}/\overline{|O_{n,m}|}^2.
\end{equation}
If the local observable operator $O_{n,m}$ has a normal distribution with a zero mean value, we have $\Gamma_{O}(\omega)=\pi/2$. The ratio $\Gamma_{O}(\omega)$ can identify the occurrence of eigenstate thermalization~\cite{PhysRevE.100.062134,PhysRevLett.125.070605,PhysRevB.102.075127,PhysRevResearch.3.043034,PhysRevE.102.062113}.

\begin{figure*}[t]
\centering
\includegraphics[scale=0.62,trim=0 0 0 0]{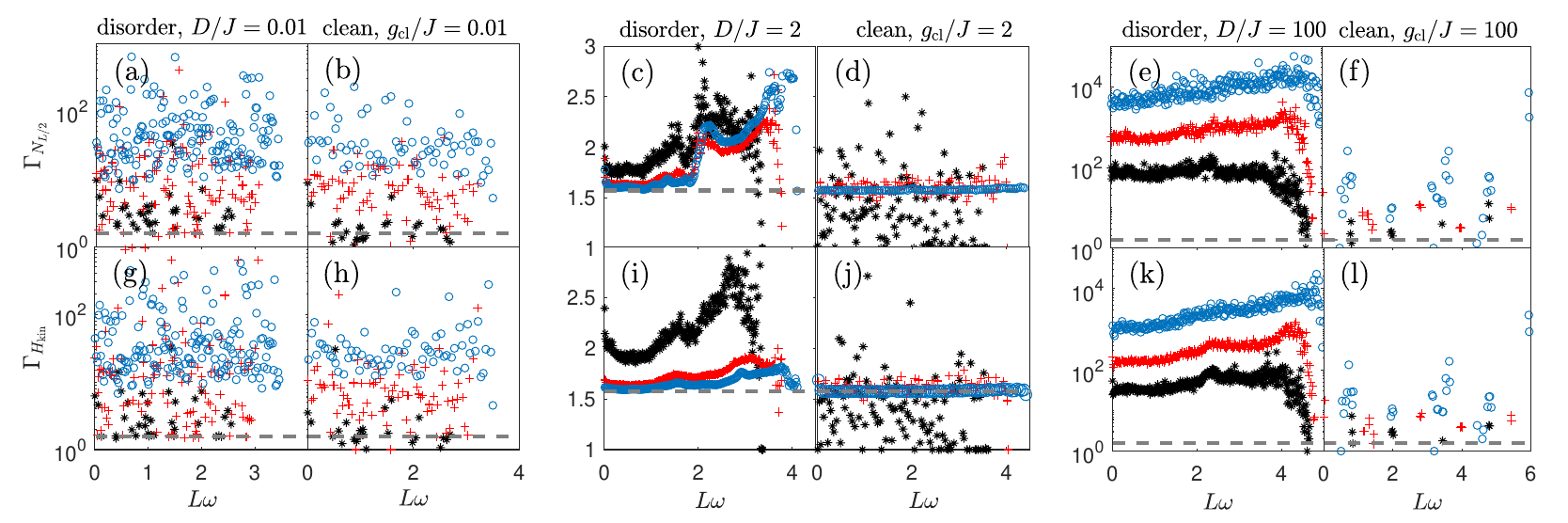}
\caption{ The ratio $\Gamma_{O}(\omega)$ for the occupancy operator $N_{L/2}$ (a)-(f) and for the kinetic operator $H_{\mathrm{kin}}$ (g)-(l) as a function of $L\omega$ in the disordered and clean JCH models.  The horizontal gray-dashed lines mark $\pi/2$ in all subgraphs. The matrix elements are computed within a small window of energy around the average spectrum $\bar{\epsilon}$ of width $0.01\omega$. The averages in $\omega$ are calculated in windows with $\delta\omega=0.002$. The black-star, red-plus, blue-circle lines correspond to $L=6$, $8$, $10$, respectively. The averaged physical quantities of the disordered JCH model are taken by $1000$, $400$, $100$ disordered samples for $L=6$, $8$, $10$, respectively.}\label{fig9}
\end{figure*}
In Fig.~\ref{fig9}, we present results of $\Gamma_{N_{L/2}}(\omega)$ and $\Gamma_{H_{\mathrm{kin}}}(\omega)$ vs. $L\omega$ in the eigenstates for the disordered and clean JCH models with the different atom-photon coupling strengths. For $D/J=0.01$ and $g_{\mathrm{cl}}/J=0.01$, one can find that the $\Gamma_{O}(\omega)$ of two observables fail to collapse, in particular in the case of large system sizes, meaning that the off-diagonal matrix elements of $N_{L/2}$ and $H_{\mathrm{kin}}$ do not obey the normal distribution. From Fig.~\ref{fig9}(c),~\ref{fig9}(d),~\ref{fig9}(i) and~\ref{fig9}(j), we find that $\Gamma_{N_{L/2}}(\omega)$ and $\Gamma_{H_{\mathrm{kin}}}(\omega)$ converge to $\pi/2$ with increasing system size for the ergodic regime of both the disordered ($D/J=2$) and clean ($g_{\mathrm{cl}}/J=2$) systems. We consider that the ratios at small $\omega$ regime have a value close to $\pi/2$. As for $D/J=100$ in Fig.~\ref{fig9}(e) and ~\ref{fig9}(k), we find that the behaviors of both $\Gamma_{N_{L/2}}(\omega)$ and $\Gamma_{H_{\mathrm{kin}}}(\omega)$ depend on the system size and do not follow a normal distribution. In addition, the clean system $g_{\mathrm{cl}}/J=100$ (Fig.~\ref{fig9}(f) and ~\ref{fig9}(l)) exhibits a similar behavior with the case of $g_{\mathrm{cl}}/J=0.01$. Neither of these cases exhibits a normal distribution, and the functions about $L\omega$ are not smooth.

From the analysis above, it is convinced that the region exhibiting MBL-like behavior does not conform to the ETH, and the ergodic region meets ETH. However, the MBL region remarkably violates ETH and its behavior is consistent with that of integrable systems~\cite{PhysRevE.100.062134,PhysRevLett.125.070605,PhysRevB.102.075127,PhysRevE.102.062113}.

\section{CONCLUSION}\label{V}
In this paper, we investigate the behavior of 1D disordered and clean JCH systems, focusing on their quasi-MBL, ergodic, and MBL phases. We also explore the similarities and differences between quasi-MBL and MBL phases. Regarding the ergodic phases, we observe that disorder has minimal impact on the system behavior at the regime that disorder strength is not strong enough. However, for strong disorder strength, the system exhibit a MBL phase, with same phenomena shown in other disordered systems.
Furthermore, we also find that the JCH model in the MBL phase displays the non-thermalization behaviors, being reminiscent of 1D integrable spin-$1/2$ system. The quasi-MBL phases also deviate the ETH. However, due to the presence of numerous quasi-degenerate energy levels, the matrix element behaviors of observables exhibit distinct characteristics compared to the conventional MBL phase, with a relatively discrete distribution.
In summary, through a comprehensive analysis of 1D disordered and clean JCH systems, we have provided insights into the impact of disorder on MBL and thermalization phenomena in these systems.

\section{acknowledgments}
This work was supported by National Natural Science Foundation of China (Grants No. 11874190, No. 61835013 and No. 12047501) and National Key R\&D Program of China under grants No. 2022YFA1404203. Support was also provided by Supercomputing Center of Lanzhou University.

\section{APPENDIX A: the effect of disorder samples}
Here, in order to explain that different disordered samples make no difference on the system results, we compare the differences between the physical quantities under multiple disordered samples.

\begin{figure}[h]
\centering
\includegraphics[width=9.2cm]{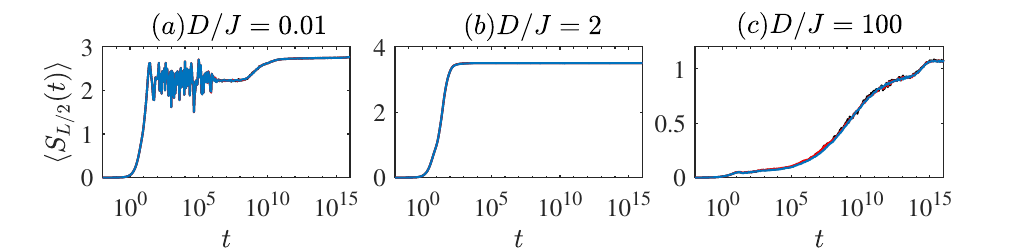}
\caption{ The average half-chain EE $\langle S_{L/2}(t)\rangle$ versus time $t$ for the disordered JCH model with different disorder strengths. Three colors represent three disordered samples [$400$ (black), $1000$ (red), $2000$ (blue)] for sizes $L=8$. }\label{fig20}
\end{figure}

From Fig.~\ref{fig20}, we can find that as the number of disorder samples increases, the behavior of the average half-chain EE $\langle S_{L/2}(t)\rangle$ becomes progressively smoother over time $t$. However, it is important to note that different disordered samples do not impact the oscillation region when the disorder strength is $D/J=0.01$. In essence, the presence of additional disordered samples does not affect the behavior of the half-chain EE, except for its fluctuations.

\begin{figure}[h]
\centering
\includegraphics[width=9.45cm]{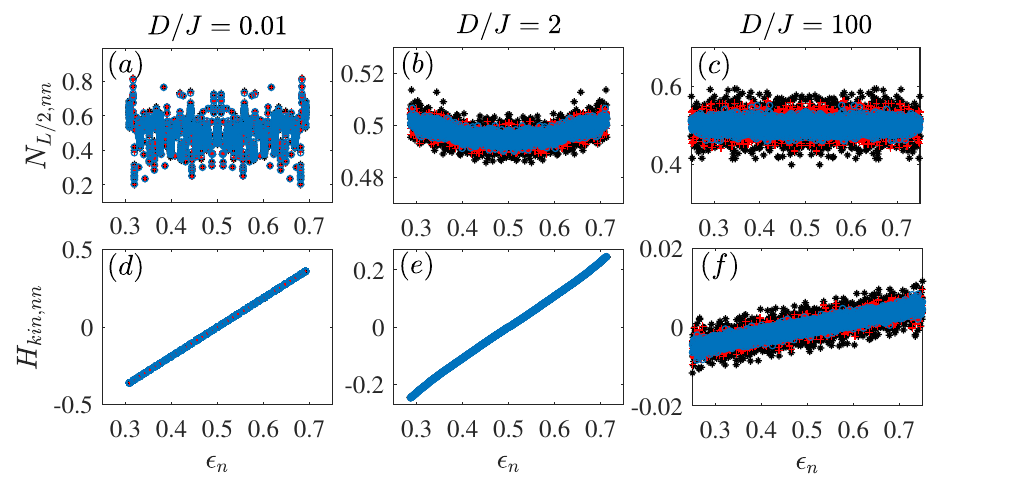}
\caption{ The diagonal matrix elements of $N_{L/2}$ (a)-(c) and $H_{\mathrm{kin}}$ (d)-(f) as a function of the energy density $\epsilon_n$ with different $D/J$ for the disordered JCH model. The black-star, red-plus, blue-circle lines correspond to $400$, $1000$, $2000$ disordered samples, respectively. The system size is chosen as $L=8$.}\label{fig21}
\end{figure}

It can be seen from Fig.~\ref{fig21} that when the disorder strength $D/J=0.01$, the disorder samples have  minimal impact on the diagonal elements of the observables. When the disorder strength $D/J=2$, with the increase of the number of disorder samples, the fluctuations of the observable $N_{L/2}$ decreases gradually. However, the observable $H_{\mathrm{kin}}$ remains unchanged since there are negligible fluctuations in $H_{\mathrm{kin}}$ as a function of the energy density $\epsilon_n$. In addition, when the disorder strength $D/J=100$, both the fluctuations of the two observables decrease with the increase of the disordered samples. On the whole, the fluctuations of the diagonal elements of the observable $H_{\mathrm{kin}}$ are smaller than that of the observable $N_{L/2}$.

\begin{figure}[h]
\centering
\includegraphics[width=9.5cm]{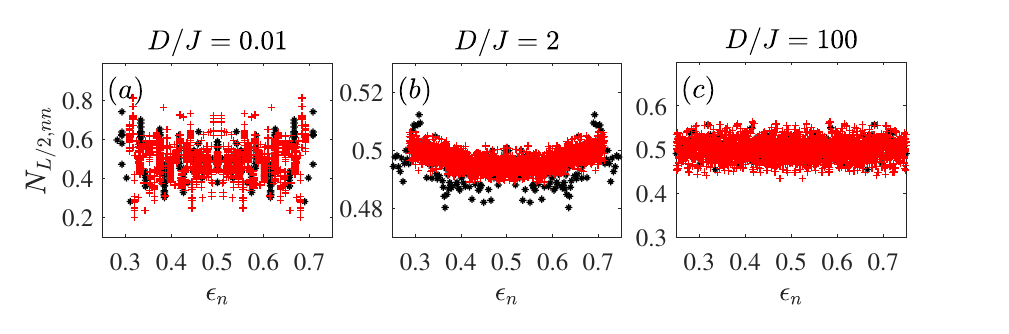}
\caption{ The diagonal matrix elements of $N_{L/2}$ as a function of the energy density $\epsilon_n$ with different $D/J$ for the disordered JCH model. The black-star, red-plus lines correspond to the system size $L=6$, $8$, respectively. The disordered samples is chosen as $1000$.}\label{fig22}
\end{figure}

By choosing the same disordered samples in Fig.~\ref{fig22}, one can easily find whether the changes of the diagonal element with sizes satisfy ETH under different disorder strengths.

\begin{figure}[h]
\centering
\includegraphics[width=8.8cm]{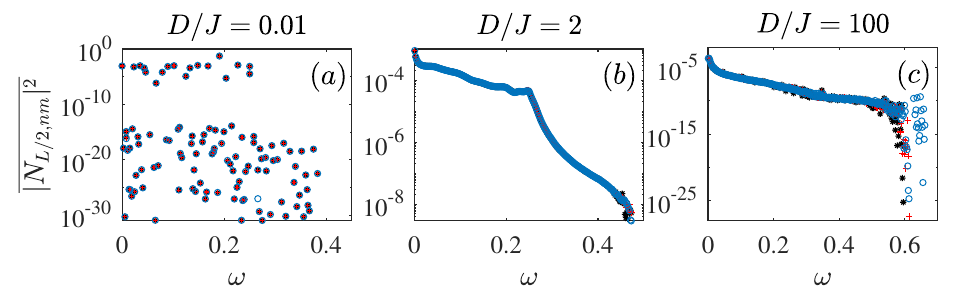}
\caption{ Coarse-grained averages of $N_{L/2}$ as a function of $\omega$ with different $D/J$ for the disordered JCH model. The matrix elements are computed within a small window of energies around the average spectrum $\bar{\epsilon}$ of width $0.01\omega$. The averages in $\omega$ are calculated in windows with $\delta\omega=0.002$.  The black-star, red-plus, blue-circle lines correspond to $400$, $1000$, $2000$ disordered samples, respectively. The system size is  $L=8$.}\label{fig23}
\end{figure}

Regarding the off-diagonal elements of the observables, the number of disordered samples also hardly affect the case of $D/J=0.01$. However, for the other two cases, increasing the number of disordered samples leads to a reduction in fluctuations, resulting in smoother functions, as shown in Fig.~\ref{fig23}. The influence of the  disordered samples on the observable $H_{\mathrm{kin}}$ (not show here) has the same behaviors as that of $N_{L/2}$.

\section{APPENDIX B: the analysis of chiral symmetry}
 In the clean case, the chiral symmetry exists only in the case of even excitations for the antisymmetric subspace.
 Let us prove the reason for this phenomenon below.
First, we consider the case where the number of excitations is even ($N\in \mathrm{even}$). The number of photons (atoms) defining the sum of odd lattice points and even lattice points is $\sum^L_{i\in \mathrm{even}}(n_i^{\mathrm{c}}+n_{i+1}^{\mathrm{c}})=N_{\mathrm{e}}^{\mathrm{c}}+N_{\mathrm{o}}^{\mathrm{c}}=N^{\mathrm{c}}$ ($\sum^L_{i\in \mathrm{even}}(n_i^{\mathrm{a}}+n_{i+1}^{\mathrm{a}})=N_{\mathrm{e}}^{\mathrm{a}}+N_{\mathrm{o}}^{\mathrm{a}}=N^{\mathrm{a}}$),
where subscript ``$\mathrm{o}$" represents odd lattice points and subscript ``$\mathrm{e}$" represents even lattice points.
At the same time, we also define the following two quantities:
$N_{\mathrm{e}}^{\mathrm{c}}+N_{\mathrm{o}}^{\mathrm{a}}=N_{1}$, $N_{\mathrm{o}}^{\mathrm{c}}+N_{\mathrm{e}}^{\mathrm{a}}=N_{2}$.

For $N_{1}+N_{2}=N\in \mathrm{even}$, then
\begin{equation}
\nonumber
\begin{cases}
N_{1} \in \mathrm{even},\\
N_{2} \in \mathrm{even},
\end{cases}
\ \ \mathrm{or} \ \
\begin{cases}
N_{1}\in \mathrm{odd},\\
N_{2}\in \mathrm{odd}.
\end{cases}
\end{equation}


When $N_{1}\in \mathrm{even}$, $N_{\mathrm{e}}^{\mathrm{c}}\in \mathrm{odd}$ and $N_{\mathrm{o}}^{\mathrm{a}}\in \mathrm{odd}$,
since the relation $L=2N$, the number of atomic ground states is $N-N_{\mathrm{o}}^{\mathrm{a}}$ (odd) for the odd number of the lattice sites. Thus, $\Gamma|\mathbf{n}\rangle=e^{i\pi N_{\mathrm{e}}^{\mathrm{c}}}(-1)^{N-N_{\mathrm{o}}^{\mathrm{a}}} \prod_{i}|n_i^{\mathrm{c}},n_i^{\mathrm{a}}\rangle_i=|\mathbf{n}\rangle$.
When $N_{1}\in \mathrm{even}$, $N_{\mathrm{e}}^{\mathrm{c}}\in \mathrm{even}$ and $N_{\mathrm{o}}^{\mathrm{a}}\in \mathrm{even}$, we can also get
$\Gamma|\mathbf{n}\rangle=|\mathbf{n}\rangle$.

For the reflective symmetric state $P|\mathbf{n}\rangle=|\mathbf{n}^{\prime}\rangle=|n_L^{\mathrm{c}},n_L^{\mathrm{a}}\rangle_1\otimes|n_{L-1}^{\mathrm{c}},n_{L-1}^{\mathrm{a}}\rangle_{2}\otimes\cdots\otimes|n_{1}^{\mathrm{c}},n_1^{\mathrm{a}}\rangle_{L}$ of state $|\mathbf{n}\rangle$, the corresponding quantity $N_{\mathrm{e}}^{\mathrm{c}\prime}+N_{\mathrm{o}}^{\mathrm{a}\prime}=N^{\prime}_{1}=N_{2}\in \mathrm{even}$, the same result can be obtained $\Gamma|\mathbf{n}^{\prime}\rangle=|\mathbf{n}^{\prime}\rangle$

When $N_{1}\in \mathrm{odd}$, $N_{\mathrm{e}}^{\mathrm{c}}\in \mathrm{even}$ and $N_{\mathrm{o}}^{\mathrm{a}}\in \mathrm{odd}$,
the number of atomic ground states is $N-N_{\mathrm{o}}^{\mathrm{a}}$ (odd) for the odd number of the lattice sites. Thus, $\Gamma|\mathbf{n}\rangle=e^{i\pi N_{\mathrm{e}}^{\mathrm{c}}}(-1)^{N-N_{\mathrm{o}}^{\mathrm{a}}} \prod_{i}|n_i,e(g)\rangle_i=-|\mathbf{n}\rangle$.
When $N_{1}\in \mathrm{odd}$, we can also get $N_{\mathrm{e}}^{\mathrm{c}}\in \mathrm{odd}$ and $N_{\mathrm{o}}^{\mathrm{a}}\in \mathrm{even}$,
$\Gamma|\mathbf{n}\rangle=-|\mathbf{n}\rangle$.

The quantity $N^{\prime}_{1}=N_{2}\in \mathrm{odd}$ for the reflective symmetric state $P|\mathbf{n}\rangle=|\mathbf{n}^{\prime}\rangle$ of state $|\mathbf{n}\rangle$, the same result can be obtained $\Gamma|\mathbf{n}^{\prime}\rangle=-|\mathbf{n}^{\prime}\rangle$

So, for a eigenstate $|n\rangle=\Sigma_i^{\mathcal{D}}\psi_i|\mathbf{n}\rangle_i$, there is
\begin{equation}
\nonumber
\begin{split}
P\Gamma|n\rangle=P\Sigma_i^{\mathcal{D}}(\pm)\psi_i|\mathbf{n}\rangle_i=\Sigma_i^{\mathcal{D}}(\pm)\psi_i|\mathbf{n}^{\prime}\rangle_i,\\
\Gamma P|n\rangle=\Gamma\Sigma_i^{\mathcal{D}}\psi_i|\mathbf{n}^{\prime}\rangle_i=\Sigma_i^{\mathcal{D}}(\pm)\psi_i|\mathbf{n}^{\prime}\rangle_i.
\end{split}
\end{equation}

This means that the two operators are commutative $[P, \Gamma]=0$ and have common eigenstates and eigenvalues.

On the other hands, for $N\in \mathrm{odd}$,then
\begin{equation}
\nonumber
\begin{cases}
N_{1}\in \mathrm{even},\\
N_{2}\in \mathrm{odd},
\end{cases}
\ \ \mathrm{or} \ \
\begin{cases}
N_{1}\in \mathrm{odd},\\
N_{2}\in \mathrm{even}.
\end{cases}
\end{equation}

When $N_{1}\in \mathrm{even}$, $N_{\mathrm{e}}^{\mathrm{c}}\in \mathrm{odd}$ and $N_{\mathrm{o}}^{\mathrm{a}}\in \mathrm{odd}$,
since the relation $L=2N$, the number of atomic ground states is $N-N_{\mathrm{o}}^{\mathrm{a}}$ (even) for the odd number of the lattice sites. Thus, $\Gamma|\mathbf{n}\rangle=e^{i\pi N_{\mathrm{e}}^{\mathrm{c}}}(-1)^{N-N_{\mathrm{o}}^{\mathrm{a}}} \prod_{i}|n_i^{\mathrm{c}},n_i^{\mathrm{a}}\rangle_i=-|\mathbf{n}\rangle$.
When $N_{1}\in \mathrm{even}$, $N_{\mathrm{e}}^{\mathrm{c}}\in \mathrm{even}$ and $N_{\mathrm{o}}^{\mathrm{a}}\in \mathrm{even}$, we can also get
$\Gamma|\mathbf{n}\rangle=-|\mathbf{n}\rangle$.

The quantity $N^{\prime}_{1}=N_{2}\in \mathrm{odd}$ for the reflective symmetric state $|\mathbf{n}^{\prime}\rangle$ of state $|\mathbf{n}\rangle$, as for $N_{\mathrm{e}}^{\mathrm{c}}\in \mathrm{even}$ and $N_{\mathrm{o}}^{\mathrm{a}}\in \mathrm{odd}$,
the number of atomic ground states is $N-N_{\mathrm{o}}^{\mathrm{a}}$ (even) for the odd number of the lattice sites. Thus, $\Gamma|\mathbf{n}\rangle=e^{i\pi N_{\mathrm{e}}^{\mathrm{c}}}(-1)^{N-N_{\mathrm{o}}^{\mathrm{a}}} \prod_{i}|n_i^{\mathrm{a}},n_i^{\mathrm{a}}\rangle_i=|\mathbf{n}\rangle$.
While for $N_{\mathrm{e}}^{\mathrm{c}}\in \mathrm{odd}$ and $N_{\mathrm{o}}^{\mathrm{a}}\in \mathrm{even}$. One also can be obtained $\Gamma|\mathbf{n}^{\prime}\rangle=|\mathbf{n}^{\prime}\rangle$.

When $N_{1}\in \mathrm{odd}$, we can also get
$\Gamma|\mathbf{n}\rangle=|\mathbf{n}\rangle$.
The quantity $N^{\prime}_{1}=N_{2}\in \mathrm{odd}$ for the reflective symmetric state $P|\mathbf{n}\rangle=|\mathbf{n}\prime\rangle$ of state $|\mathbf{n}\rangle$, the same can be obtained $\Gamma|\mathbf{n}^{\prime}\rangle=-|\mathbf{n}^{\prime}\rangle$

So, for a eigenstate $|n\rangle=\Sigma_i^{\mathcal{D}}\psi_i|\mathbf{n}\rangle_i$, there is
\begin{equation}
\nonumber
\begin{split}
P\Gamma|n\rangle=P\Sigma_i^{\mathcal{D}}(\pm)\psi_i|\mathbf{n}\rangle_i=\Sigma_i^{\mathcal{D}}(\pm)\psi_i|\mathbf{n}^{\prime}\rangle_i,\\
\Gamma P|n\rangle=\Gamma\Sigma_i^{\mathcal{D}}\psi_i|\mathbf{n}^{\prime}\rangle_i=\Sigma_i^{\mathcal{D}}(\mp)\psi_i|\mathbf{n}^{\prime}\rangle_i.
\end{split}
\end{equation}
This means that the two operators are commutative $[P, \Gamma]\neq0$ and without common eigenstates and eigenvalues.


\begin{thebibliography}{52}%
\makeatletter
\providecommand \@ifxundefined [1]{%
 \@ifx{#1\undefined}
}%
\providecommand \@ifnum [1]{%
 \ifnum #1\expandafter \@firstoftwo
 \else \expandafter \@secondoftwo
 \fi
}%
\providecommand \@ifx [1]{%
 \ifx #1\expandafter \@firstoftwo
 \else \expandafter \@secondoftwo
 \fi
}%
\providecommand \natexlab [1]{#1}%
\providecommand \enquote  [1]{``#1''}%
\providecommand \bibnamefont  [1]{#1}%
\providecommand \bibfnamefont [1]{#1}%
\providecommand \citenamefont [1]{#1}%
\providecommand \href@noop [0]{\@secondoftwo}%
\providecommand \href [0]{\begingroup \@sanitize@url \@href}%
\providecommand \@href[1]{\@@startlink{#1}\@@href}%
\providecommand \@@href[1]{\endgroup#1\@@endlink}%
\providecommand \@sanitize@url [0]{\catcode `\\12\catcode `\$12\catcode
  `\&12\catcode `\#12\catcode `\^12\catcode `\_12\catcode `\%12\relax}%
\providecommand \@@startlink[1]{}%
\providecommand \@@endlink[0]{}%
\providecommand \url  [0]{\begingroup\@sanitize@url \@url }%
\providecommand \@url [1]{\endgroup\@href {#1}{\urlprefix }}%
\providecommand \urlprefix  [0]{URL }%
\providecommand \Eprint [0]{\href }%
\providecommand \doibase [0]{https://doi.org/}%
\providecommand \selectlanguage [0]{\@gobble}%
\providecommand \bibinfo  [0]{\@secondoftwo}%
\providecommand \bibfield  [0]{\@secondoftwo}%
\providecommand \translation [1]{[#1]}%
\providecommand \BibitemOpen [0]{}%
\providecommand \bibitemStop [0]{}%
\providecommand \bibitemNoStop [0]{.\EOS\space}%
\providecommand \EOS [0]{\spacefactor3000\relax}%
\providecommand \BibitemShut  [1]{\csname bibitem#1\endcsname}%
\let\auto@bib@innerbib\@empty
\bibitem [{\citenamefont {Rigol}\ \emph {et~al.}(2008)\citenamefont {Rigol},
  \citenamefont {Dunjko},\ and\ \citenamefont
  {Olshanii}}]{rigol2008thermalization}%
  \BibitemOpen
  \bibfield  {author} {\bibinfo {author} {\bibfnamefont {M.}~\bibnamefont
  {Rigol}}, \bibinfo {author} {\bibfnamefont {V.}~\bibnamefont {Dunjko}},\ and\
  \bibinfo {author} {\bibfnamefont {M.}~\bibnamefont {Olshanii}},\ }\bibfield
  {title} {\bibinfo {title} {Thermalization and its mechanism for generic
  isolated quantum systems},\ }\href {https://doi.org/10.1038/nature06838}
  {\bibfield  {journal} {\bibinfo  {journal} {Nature}\ }\textbf {\bibinfo
  {volume} {452}},\ \bibinfo {pages} {854} (\bibinfo {year}
  {2008})}\BibitemShut {NoStop}%
\bibitem [{\citenamefont {Rigol}(2009{\natexlab{a}})}]{PhysRevLett.103.100403}%
  \BibitemOpen
  \bibfield  {author} {\bibinfo {author} {\bibfnamefont {M.}~\bibnamefont
  {Rigol}},\ }\bibfield  {title} {\bibinfo {title} {Breakdown of thermalization
  in finite one-dimensional systems},\ }\href
  {https://doi.org/10.1103/PhysRevLett.103.100403} {\bibfield  {journal}
  {\bibinfo  {journal} {Phys. Rev. Lett.}\ }\textbf {\bibinfo {volume} {103}},\
  \bibinfo {pages} {100403} (\bibinfo {year} {2009}{\natexlab{a}})}\BibitemShut
  {NoStop}%
\bibitem [{\citenamefont {Sorg}\ \emph {et~al.}(2014)\citenamefont {Sorg},
  \citenamefont {Vidmar}, \citenamefont {Pollet},\ and\ \citenamefont
  {Heidrich-Meisner}}]{PhysRevA.90.033606}%
  \BibitemOpen
  \bibfield  {author} {\bibinfo {author} {\bibfnamefont {S.}~\bibnamefont
  {Sorg}}, \bibinfo {author} {\bibfnamefont {L.}~\bibnamefont {Vidmar}},
  \bibinfo {author} {\bibfnamefont {L.}~\bibnamefont {Pollet}},\ and\ \bibinfo
  {author} {\bibfnamefont {F.}~\bibnamefont {Heidrich-Meisner}},\ }\bibfield
  {title} {\bibinfo {title} {Relaxation and thermalization in the
  one-dimensional bose-hubbard model: A case study for the interaction quantum
  quench from the atomic limit},\ }\href
  {https://doi.org/10.1103/PhysRevA.90.033606} {\bibfield  {journal} {\bibinfo
  {journal} {Phys. Rev. A}\ }\textbf {\bibinfo {volume} {90}},\ \bibinfo
  {pages} {033606} (\bibinfo {year} {2014})}\BibitemShut {NoStop}%
\bibitem [{\citenamefont {Rigol}(2009{\natexlab{b}})}]{PhysRevA.80.053607}%
  \BibitemOpen
  \bibfield  {author} {\bibinfo {author} {\bibfnamefont {M.}~\bibnamefont
  {Rigol}},\ }\bibfield  {title} {\bibinfo {title} {Quantum quenches and
  thermalization in one-dimensional fermionic systems},\ }\href
  {https://doi.org/10.1103/PhysRevA.80.053607} {\bibfield  {journal} {\bibinfo
  {journal} {Phys. Rev. A}\ }\textbf {\bibinfo {volume} {80}},\ \bibinfo
  {pages} {053607} (\bibinfo {year} {2009}{\natexlab{b}})}\BibitemShut
  {NoStop}%
\bibitem [{\citenamefont {Jansen}\ \emph {et~al.}(2019)\citenamefont {Jansen},
  \citenamefont {Stolpp}, \citenamefont {Vidmar},\ and\ \citenamefont
  {Heidrich-Meisner}}]{PhysRevB.99.155130}%
  \BibitemOpen
  \bibfield  {author} {\bibinfo {author} {\bibfnamefont {D.}~\bibnamefont
  {Jansen}}, \bibinfo {author} {\bibfnamefont {J.}~\bibnamefont {Stolpp}},
  \bibinfo {author} {\bibfnamefont {L.}~\bibnamefont {Vidmar}},\ and\ \bibinfo
  {author} {\bibfnamefont {F.}~\bibnamefont {Heidrich-Meisner}},\ }\bibfield
  {title} {\bibinfo {title} {Eigenstate thermalization and quantum chaos in the
  holstein polaron model},\ }\href {https://doi.org/10.1103/PhysRevB.99.155130}
  {\bibfield  {journal} {\bibinfo  {journal} {Phys. Rev. B}\ }\textbf {\bibinfo
  {volume} {99}},\ \bibinfo {pages} {155130} (\bibinfo {year}
  {2019})}\BibitemShut {NoStop}%
\bibitem [{\citenamefont {Mondaini}\ \emph {et~al.}(2016)\citenamefont
  {Mondaini}, \citenamefont {Fratus}, \citenamefont {Srednicki},\ and\
  \citenamefont {Rigol}}]{PhysRevE.93.032104}%
  \BibitemOpen
  \bibfield  {author} {\bibinfo {author} {\bibfnamefont {R.}~\bibnamefont
  {Mondaini}}, \bibinfo {author} {\bibfnamefont {K.~R.}\ \bibnamefont
  {Fratus}}, \bibinfo {author} {\bibfnamefont {M.}~\bibnamefont {Srednicki}},\
  and\ \bibinfo {author} {\bibfnamefont {M.}~\bibnamefont {Rigol}},\ }\bibfield
   {title} {\bibinfo {title} {Eigenstate thermalization in the two-dimensional
  transverse field ising model},\ }\href
  {https://doi.org/10.1103/PhysRevE.93.032104} {\bibfield  {journal} {\bibinfo
  {journal} {Phys. Rev. E}\ }\textbf {\bibinfo {volume} {93}},\ \bibinfo
  {pages} {032104} (\bibinfo {year} {2016})}\BibitemShut {NoStop}%
\bibitem [{\citenamefont {Steinigeweg}\ \emph {et~al.}(2013)\citenamefont
  {Steinigeweg}, \citenamefont {Herbrych},\ and\ \citenamefont
  {Prelov\ifmmode~\check{s}\else \v{s}\fi{}ek}}]{PhysRevE.87.012118}%
  \BibitemOpen
  \bibfield  {author} {\bibinfo {author} {\bibfnamefont {R.}~\bibnamefont
  {Steinigeweg}}, \bibinfo {author} {\bibfnamefont {J.}~\bibnamefont
  {Herbrych}},\ and\ \bibinfo {author} {\bibfnamefont {P.}~\bibnamefont
  {Prelov\ifmmode~\check{s}\else \v{s}\fi{}ek}},\ }\bibfield  {title} {\bibinfo
  {title} {Eigenstate thermalization within isolated spin-chain systems},\
  }\href {https://doi.org/10.1103/PhysRevE.87.012118} {\bibfield  {journal}
  {\bibinfo  {journal} {Phys. Rev. E}\ }\textbf {\bibinfo {volume} {87}},\
  \bibinfo {pages} {012118} (\bibinfo {year} {2013})}\BibitemShut {NoStop}%
\bibitem [{\citenamefont {Beugeling}\ \emph {et~al.}(2014)\citenamefont
  {Beugeling}, \citenamefont {Moessner},\ and\ \citenamefont
  {Haque}}]{PhysRevE.89.042112}%
  \BibitemOpen
  \bibfield  {author} {\bibinfo {author} {\bibfnamefont {W.}~\bibnamefont
  {Beugeling}}, \bibinfo {author} {\bibfnamefont {R.}~\bibnamefont
  {Moessner}},\ and\ \bibinfo {author} {\bibfnamefont {M.}~\bibnamefont
  {Haque}},\ }\bibfield  {title} {\bibinfo {title} {Finite-size scaling of
  eigenstate thermalization},\ }\href
  {https://doi.org/10.1103/PhysRevE.89.042112} {\bibfield  {journal} {\bibinfo
  {journal} {Phys. Rev. E}\ }\textbf {\bibinfo {volume} {89}},\ \bibinfo
  {pages} {042112} (\bibinfo {year} {2014})}\BibitemShut {NoStop}%
\bibitem [{\citenamefont {Kim}\ \emph {et~al.}(2014)\citenamefont {Kim},
  \citenamefont {Ikeda},\ and\ \citenamefont {Huse}}]{PhysRevE.90.052105}%
  \BibitemOpen
  \bibfield  {author} {\bibinfo {author} {\bibfnamefont {H.}~\bibnamefont
  {Kim}}, \bibinfo {author} {\bibfnamefont {T.~N.}\ \bibnamefont {Ikeda}},\
  and\ \bibinfo {author} {\bibfnamefont {D.~A.}\ \bibnamefont {Huse}},\
  }\bibfield  {title} {\bibinfo {title} {Testing whether all eigenstates obey
  the eigenstate thermalization hypothesis},\ }\href
  {https://doi.org/10.1103/PhysRevE.90.052105} {\bibfield  {journal} {\bibinfo
  {journal} {Phys. Rev. E}\ }\textbf {\bibinfo {volume} {90}},\ \bibinfo
  {pages} {052105} (\bibinfo {year} {2014})}\BibitemShut {NoStop}%
\bibitem [{\citenamefont {Yoshizawa}\ \emph {et~al.}(2018)\citenamefont
  {Yoshizawa}, \citenamefont {Iyoda},\ and\ \citenamefont
  {Sagawa}}]{PhysRevLett.120.200604}%
  \BibitemOpen
  \bibfield  {author} {\bibinfo {author} {\bibfnamefont {T.}~\bibnamefont
  {Yoshizawa}}, \bibinfo {author} {\bibfnamefont {E.}~\bibnamefont {Iyoda}},\
  and\ \bibinfo {author} {\bibfnamefont {T.}~\bibnamefont {Sagawa}},\
  }\bibfield  {title} {\bibinfo {title} {Numerical large deviation analysis of
  the eigenstate thermalization hypothesis},\ }\href
  {https://doi.org/10.1103/PhysRevLett.120.200604} {\bibfield  {journal}
  {\bibinfo  {journal} {Phys. Rev. Lett.}\ }\textbf {\bibinfo {volume} {120}},\
  \bibinfo {pages} {200604} (\bibinfo {year} {2018})}\BibitemShut {NoStop}%
\bibitem [{\citenamefont {Khatami}\ \emph {et~al.}(2013)\citenamefont
  {Khatami}, \citenamefont {Pupillo}, \citenamefont {Srednicki},\ and\
  \citenamefont {Rigol}}]{PhysRevLett.111.050403}%
  \BibitemOpen
  \bibfield  {author} {\bibinfo {author} {\bibfnamefont {E.}~\bibnamefont
  {Khatami}}, \bibinfo {author} {\bibfnamefont {G.}~\bibnamefont {Pupillo}},
  \bibinfo {author} {\bibfnamefont {M.}~\bibnamefont {Srednicki}},\ and\
  \bibinfo {author} {\bibfnamefont {M.}~\bibnamefont {Rigol}},\ }\bibfield
  {title} {\bibinfo {title} {Fluctuation-dissipation theorem in an isolated
  system of quantum dipolar bosons after a quench},\ }\href
  {https://doi.org/10.1103/PhysRevLett.111.050403} {\bibfield  {journal}
  {\bibinfo  {journal} {Phys. Rev. Lett.}\ }\textbf {\bibinfo {volume} {111}},\
  \bibinfo {pages} {050403} (\bibinfo {year} {2013})}\BibitemShut {NoStop}%
\bibitem [{\citenamefont {Mondaini}\ and\ \citenamefont
  {Rigol}(2017)}]{PhysRevE.96.012157}%
  \BibitemOpen
  \bibfield  {author} {\bibinfo {author} {\bibfnamefont {R.}~\bibnamefont
  {Mondaini}}\ and\ \bibinfo {author} {\bibfnamefont {M.}~\bibnamefont
  {Rigol}},\ }\bibfield  {title} {\bibinfo {title} {Eigenstate thermalization
  in the two-dimensional transverse field ising model. ii. off-diagonal matrix
  elements of observables},\ }\href
  {https://doi.org/10.1103/PhysRevE.96.012157} {\bibfield  {journal} {\bibinfo
  {journal} {Phys. Rev. E}\ }\textbf {\bibinfo {volume} {96}},\ \bibinfo
  {pages} {012157} (\bibinfo {year} {2017})}\BibitemShut {NoStop}%
\bibitem [{\citenamefont {Santos}\ and\ \citenamefont
  {Rigol}(2010)}]{PhysRevE.82.031130}%
  \BibitemOpen
  \bibfield  {author} {\bibinfo {author} {\bibfnamefont {L.~F.}\ \bibnamefont
  {Santos}}\ and\ \bibinfo {author} {\bibfnamefont {M.}~\bibnamefont {Rigol}},\
  }\bibfield  {title} {\bibinfo {title} {Localization and the effects of
  symmetries in the thermalization properties of one-dimensional quantum
  systems},\ }\href {https://doi.org/10.1103/PhysRevE.82.031130} {\bibfield
  {journal} {\bibinfo  {journal} {Phys. Rev. E}\ }\textbf {\bibinfo {volume}
  {82}},\ \bibinfo {pages} {031130} (\bibinfo {year} {2010})}\BibitemShut
  {NoStop}%
\bibitem [{\citenamefont {Steinigeweg}\ \emph {et~al.}(2014)\citenamefont
  {Steinigeweg}, \citenamefont {Khodja}, \citenamefont {Niemeyer},
  \citenamefont {Gogolin},\ and\ \citenamefont
  {Gemmer}}]{PhysRevLett.112.130403}%
  \BibitemOpen
  \bibfield  {author} {\bibinfo {author} {\bibfnamefont {R.}~\bibnamefont
  {Steinigeweg}}, \bibinfo {author} {\bibfnamefont {A.}~\bibnamefont {Khodja}},
  \bibinfo {author} {\bibfnamefont {H.}~\bibnamefont {Niemeyer}}, \bibinfo
  {author} {\bibfnamefont {C.}~\bibnamefont {Gogolin}},\ and\ \bibinfo {author}
  {\bibfnamefont {J.}~\bibnamefont {Gemmer}},\ }\bibfield  {title} {\bibinfo
  {title} {Pushing the limits of the eigenstate thermalization hypothesis
  towards mesoscopic quantum systems},\ }\href
  {https://doi.org/10.1103/PhysRevLett.112.130403} {\bibfield  {journal}
  {\bibinfo  {journal} {Phys. Rev. Lett.}\ }\textbf {\bibinfo {volume} {112}},\
  \bibinfo {pages} {130403} (\bibinfo {year} {2014})}\BibitemShut {NoStop}%
\bibitem [{\citenamefont {Khaymovich}\ \emph {et~al.}(2019)\citenamefont
  {Khaymovich}, \citenamefont {Haque},\ and\ \citenamefont
  {McClarty}}]{PhysRevLett.122.070601}%
  \BibitemOpen
  \bibfield  {author} {\bibinfo {author} {\bibfnamefont {I.~M.}\ \bibnamefont
  {Khaymovich}}, \bibinfo {author} {\bibfnamefont {M.}~\bibnamefont {Haque}},\
  and\ \bibinfo {author} {\bibfnamefont {P.~A.}\ \bibnamefont {McClarty}},\
  }\bibfield  {title} {\bibinfo {title} {Eigenstate thermalization, random
  matrix theory, and behemoths},\ }\href
  {https://doi.org/10.1103/PhysRevLett.122.070601} {\bibfield  {journal}
  {\bibinfo  {journal} {Phys. Rev. Lett.}\ }\textbf {\bibinfo {volume} {122}},\
  \bibinfo {pages} {070601} (\bibinfo {year} {2019})}\BibitemShut {NoStop}%
\bibitem [{\citenamefont {Mierzejewski}\ and\ \citenamefont
  {Vidmar}(2020)}]{PhysRevLett.124.040603}%
  \BibitemOpen
  \bibfield  {author} {\bibinfo {author} {\bibfnamefont {M.}~\bibnamefont
  {Mierzejewski}}\ and\ \bibinfo {author} {\bibfnamefont {L.}~\bibnamefont
  {Vidmar}},\ }\bibfield  {title} {\bibinfo {title} {Quantitative impact of
  integrals of motion on the eigenstate thermalization hypothesis},\ }\href
  {https://doi.org/10.1103/PhysRevLett.124.040603} {\bibfield  {journal}
  {\bibinfo  {journal} {Phys. Rev. Lett.}\ }\textbf {\bibinfo {volume} {124}},\
  \bibinfo {pages} {040603} (\bibinfo {year} {2020})}\BibitemShut {NoStop}%
\bibitem [{\citenamefont {Bauer}\ and\ \citenamefont
  {Nayak}(2013)}]{Bauer_2013}%
  \BibitemOpen
  \bibfield  {author} {\bibinfo {author} {\bibfnamefont {B.}~\bibnamefont
  {Bauer}}\ and\ \bibinfo {author} {\bibfnamefont {C.}~\bibnamefont {Nayak}},\
  }\bibfield  {title} {\bibinfo {title} {Area laws in a many-body localized
  state and its implications for topological order},\ }\href
  {https://doi.org/10.1088/1742-5468/2013/09/P09005} {\bibfield  {journal}
  {\bibinfo  {journal} {Journal of Statistical Mechanics: Theory and
  Experiment}\ }\textbf {\bibinfo {volume} {2013}},\ \bibinfo {pages} {P09005}
  (\bibinfo {year} {2013})}\BibitemShut {NoStop}%
\bibitem [{\citenamefont {Huse}\ \emph {et~al.}(2013)\citenamefont {Huse},
  \citenamefont {Nandkishore}, \citenamefont {Oganesyan}, \citenamefont {Pal},\
  and\ \citenamefont {Sondhi}}]{PhysRevB.88.014206}%
  \BibitemOpen
  \bibfield  {author} {\bibinfo {author} {\bibfnamefont {D.~A.}\ \bibnamefont
  {Huse}}, \bibinfo {author} {\bibfnamefont {R.}~\bibnamefont {Nandkishore}},
  \bibinfo {author} {\bibfnamefont {V.}~\bibnamefont {Oganesyan}}, \bibinfo
  {author} {\bibfnamefont {A.}~\bibnamefont {Pal}},\ and\ \bibinfo {author}
  {\bibfnamefont {S.~L.}\ \bibnamefont {Sondhi}},\ }\bibfield  {title}
  {\bibinfo {title} {Localization-protected quantum order},\ }\href
  {https://doi.org/10.1103/PhysRevB.88.014206} {\bibfield  {journal} {\bibinfo
  {journal} {Phys. Rev. B}\ }\textbf {\bibinfo {volume} {88}},\ \bibinfo
  {pages} {014206} (\bibinfo {year} {2013})}\BibitemShut {NoStop}%
\bibitem [{\citenamefont {Parameswaran}\ and\ \citenamefont
  {Vasseur}(2018)}]{Parameswaran_2018}%
  \BibitemOpen
  \bibfield  {author} {\bibinfo {author} {\bibfnamefont {S.~A.}\ \bibnamefont
  {Parameswaran}}\ and\ \bibinfo {author} {\bibfnamefont {R.}~\bibnamefont
  {Vasseur}},\ }\bibfield  {title} {\bibinfo {title} {Many-body localization,
  symmetry and topology},\ }\href {https://doi.org/10.1088/1361-6633/aac9ed}
  {\bibfield  {journal} {\bibinfo  {journal} {Reports on Progress in Physics}\
  }\textbf {\bibinfo {volume} {81}},\ \bibinfo {pages} {082501} (\bibinfo
  {year} {2018})}\BibitemShut {NoStop}%
\bibitem [{\citenamefont {Chandran}\ \emph {et~al.}(2014)\citenamefont
  {Chandran}, \citenamefont {Khemani}, \citenamefont {Laumann},\ and\
  \citenamefont {Sondhi}}]{PhysRevB.89.144201}%
  \BibitemOpen
  \bibfield  {author} {\bibinfo {author} {\bibfnamefont {A.}~\bibnamefont
  {Chandran}}, \bibinfo {author} {\bibfnamefont {V.}~\bibnamefont {Khemani}},
  \bibinfo {author} {\bibfnamefont {C.~R.}\ \bibnamefont {Laumann}},\ and\
  \bibinfo {author} {\bibfnamefont {S.~L.}\ \bibnamefont {Sondhi}},\ }\bibfield
   {title} {\bibinfo {title} {Many-body localization and symmetry-protected
  topological order},\ }\href {https://doi.org/10.1103/PhysRevB.89.144201}
  {\bibfield  {journal} {\bibinfo  {journal} {Phys. Rev. B}\ }\textbf {\bibinfo
  {volume} {89}},\ \bibinfo {pages} {144201} (\bibinfo {year}
  {2014})}\BibitemShut {NoStop}%
\bibitem [{\citenamefont {Kj\"all}\ \emph {et~al.}(2014)\citenamefont
  {Kj\"all}, \citenamefont {Bardarson},\ and\ \citenamefont
  {Pollmann}}]{PhysRevLett.113.107204}%
  \BibitemOpen
  \bibfield  {author} {\bibinfo {author} {\bibfnamefont {J.~A.}\ \bibnamefont
  {Kj\"all}}, \bibinfo {author} {\bibfnamefont {J.~H.}\ \bibnamefont
  {Bardarson}},\ and\ \bibinfo {author} {\bibfnamefont {F.}~\bibnamefont
  {Pollmann}},\ }\bibfield  {title} {\bibinfo {title} {Many-body localization
  in a disordered quantum ising chain},\ }\href
  {https://doi.org/10.1103/PhysRevLett.113.107204} {\bibfield  {journal}
  {\bibinfo  {journal} {Phys. Rev. Lett.}\ }\textbf {\bibinfo {volume} {113}},\
  \bibinfo {pages} {107204} (\bibinfo {year} {2014})}\BibitemShut {NoStop}%
\bibitem [{\citenamefont {Pekker}\ \emph {et~al.}(2014)\citenamefont {Pekker},
  \citenamefont {Refael}, \citenamefont {Altman}, \citenamefont {Demler},\ and\
  \citenamefont {Oganesyan}}]{PhysRevX.4.011052}%
  \BibitemOpen
  \bibfield  {author} {\bibinfo {author} {\bibfnamefont {D.}~\bibnamefont
  {Pekker}}, \bibinfo {author} {\bibfnamefont {G.}~\bibnamefont {Refael}},
  \bibinfo {author} {\bibfnamefont {E.}~\bibnamefont {Altman}}, \bibinfo
  {author} {\bibfnamefont {E.}~\bibnamefont {Demler}},\ and\ \bibinfo {author}
  {\bibfnamefont {V.}~\bibnamefont {Oganesyan}},\ }\bibfield  {title} {\bibinfo
  {title} {Hilbert-glass transition: New universality of temperature-tuned
  many-body dynamical quantum criticality},\ }\href
  {https://doi.org/10.1103/PhysRevX.4.011052} {\bibfield  {journal} {\bibinfo
  {journal} {Phys. Rev. X}\ }\textbf {\bibinfo {volume} {4}},\ \bibinfo {pages}
  {011052} (\bibinfo {year} {2014})}\BibitemShut {NoStop}%
\bibitem [{\citenamefont {Alet}\ and\ \citenamefont
  {Laflorencie}(2018)}]{ALET2018498}%
  \BibitemOpen
  \bibfield  {author} {\bibinfo {author} {\bibfnamefont {F.}~\bibnamefont
  {Alet}}\ and\ \bibinfo {author} {\bibfnamefont {N.}~\bibnamefont
  {Laflorencie}},\ }\bibfield  {title} {\bibinfo {title} {Many-body
  localization: An introduction and selected topics},\ }\href
  {https://doi.org/https://doi.org/10.1016/j.crhy.2018.03.003} {\bibfield
  {journal} {\bibinfo  {journal} {Comptes Rendus Physique}\ }\textbf {\bibinfo
  {volume} {19}},\ \bibinfo {pages} {498} (\bibinfo {year} {2018})},\ \bibinfo
  {note} {quantum simulation / Simulation quantique}\BibitemShut {NoStop}%
\bibitem [{\citenamefont {Nandkishore}\ and\ \citenamefont
  {Huse}(2015)}]{doi:10.1146/annurev-conmatphys-031214-014726}%
  \BibitemOpen
  \bibfield  {author} {\bibinfo {author} {\bibfnamefont {R.}~\bibnamefont
  {Nandkishore}}\ and\ \bibinfo {author} {\bibfnamefont {D.~A.}\ \bibnamefont
  {Huse}},\ }\bibfield  {title} {\bibinfo {title} {Many-body localization and
  thermalization in quantum statistical mechanics},\ }\href
  {https://doi.org/10.1146/annurev-conmatphys-031214-014726} {\bibfield
  {journal} {\bibinfo  {journal} {Annual Review of Condensed Matter Physics}\
  }\textbf {\bibinfo {volume} {6}},\ \bibinfo {pages} {15} (\bibinfo {year}
  {2015})},\ \Eprint
  {https://arxiv.org/abs/https://doi.org/10.1146/annurev-conmatphys-031214-014726}
  {https://doi.org/10.1146/annurev-conmatphys-031214-014726} \BibitemShut
  {NoStop}%
\bibitem [{\citenamefont {Hartmann}\ \emph {et~al.}(2008)\citenamefont
  {Hartmann}, \citenamefont {Brandão},\ and\ \citenamefont
  {Plenio}}]{https://doi.org/10.1002/lpor.200810046}%
  \BibitemOpen
  \bibfield  {author} {\bibinfo {author} {\bibfnamefont {M.}~\bibnamefont
  {Hartmann}}, \bibinfo {author} {\bibfnamefont {F.}~\bibnamefont {Brandão}},\
  and\ \bibinfo {author} {\bibfnamefont {M.}~\bibnamefont {Plenio}},\
  }\bibfield  {title} {\bibinfo {title} {Quantum many-body phenomena in coupled
  cavity arrays},\ }\href
  {https://doi.org/https://doi.org/10.1002/lpor.200810046} {\bibfield
  {journal} {\bibinfo  {journal} {Laser \& Photonics Reviews}\ }\textbf
  {\bibinfo {volume} {2}},\ \bibinfo {pages} {527} (\bibinfo {year} {2008})},\
  \Eprint
  {https://arxiv.org/abs/https://onlinelibrary.wiley.com/doi/pdf/10.1002/lpor.200810046}
  {https://onlinelibrary.wiley.com/doi/pdf/10.1002/lpor.200810046} \BibitemShut
  {NoStop}%
\bibitem [{\citenamefont {Li}\ \emph {et~al.}(2021{\natexlab{a}})\citenamefont
  {Li}, \citenamefont {Ma}, \citenamefont {Huang}, \citenamefont {Tan},
  \citenamefont {Gu},\ and\ \citenamefont {Liu}}]{Li_2021}%
  \BibitemOpen
  \bibfield  {author} {\bibinfo {author} {\bibfnamefont {Q.}~\bibnamefont
  {Li}}, \bibinfo {author} {\bibfnamefont {J.-L.}\ \bibnamefont {Ma}}, \bibinfo
  {author} {\bibfnamefont {T.}~\bibnamefont {Huang}}, \bibinfo {author}
  {\bibfnamefont {L.}~\bibnamefont {Tan}}, \bibinfo {author} {\bibfnamefont
  {H.-Q.}\ \bibnamefont {Gu}},\ and\ \bibinfo {author} {\bibfnamefont {W.-M.}\
  \bibnamefont {Liu}},\ }\bibfield  {title} {\bibinfo {title} {Quantum quench
  dynamics of the jaynes-cummings-hubbard model with weak nearest-neighbor
  hopping},\ }\href {https://doi.org/10.1209/0295-5075/134/20007} {\bibfield
  {journal} {\bibinfo  {journal} {Europhysics Letters}\ }\textbf {\bibinfo
  {volume} {134}},\ \bibinfo {pages} {20007} (\bibinfo {year}
  {2021}{\natexlab{a}})}\BibitemShut {NoStop}%
\bibitem [{\citenamefont {Li}\ \emph {et~al.}(2021{\natexlab{b}})\citenamefont
  {Li}, \citenamefont {Ma},\ and\ \citenamefont {Tan}}]{Li_2021b}%
  \BibitemOpen
  \bibfield  {author} {\bibinfo {author} {\bibfnamefont {Q.}~\bibnamefont
  {Li}}, \bibinfo {author} {\bibfnamefont {J.-L.}\ \bibnamefont {Ma}},\ and\
  \bibinfo {author} {\bibfnamefont {L.}~\bibnamefont {Tan}},\ }\bibfield
  {title} {\bibinfo {title} {Eigenstate thermalization and quantum chaos in the
  jaynes–cummings hubbard model},\ }\href
  {https://doi.org/10.1088/1402-4896/ac267f} {\bibfield  {journal} {\bibinfo
  {journal} {Physica Scripta}\ }\textbf {\bibinfo {volume} {96}},\ \bibinfo
  {pages} {125709} (\bibinfo {year} {2021}{\natexlab{b}})}\BibitemShut
  {NoStop}%
\bibitem [{\citenamefont {Yao}\ \emph {et~al.}(2016)\citenamefont {Yao},
  \citenamefont {Laumann}, \citenamefont {Cirac}, \citenamefont {Lukin},\ and\
  \citenamefont {Moore}}]{PhysRevLett.117.240601}%
  \BibitemOpen
  \bibfield  {author} {\bibinfo {author} {\bibfnamefont {N.~Y.}\ \bibnamefont
  {Yao}}, \bibinfo {author} {\bibfnamefont {C.~R.}\ \bibnamefont {Laumann}},
  \bibinfo {author} {\bibfnamefont {J.~I.}\ \bibnamefont {Cirac}}, \bibinfo
  {author} {\bibfnamefont {M.~D.}\ \bibnamefont {Lukin}},\ and\ \bibinfo
  {author} {\bibfnamefont {J.~E.}\ \bibnamefont {Moore}},\ }\bibfield  {title}
  {\bibinfo {title} {Quasi-many-body localization in translation-invariant
  systems},\ }\href {https://doi.org/10.1103/PhysRevLett.117.240601} {\bibfield
   {journal} {\bibinfo  {journal} {Phys. Rev. Lett.}\ }\textbf {\bibinfo
  {volume} {117}},\ \bibinfo {pages} {240601} (\bibinfo {year}
  {2016})}\BibitemShut {NoStop}%
\bibitem [{\citenamefont {Mascarenhas}\ \emph {et~al.}(2012)\citenamefont
  {Mascarenhas}, \citenamefont {Heaney}, \citenamefont {Aguiar},\ and\
  \citenamefont {Santos}}]{Mascarenhas_2012}%
  \BibitemOpen
  \bibfield  {author} {\bibinfo {author} {\bibfnamefont {E.}~\bibnamefont
  {Mascarenhas}}, \bibinfo {author} {\bibfnamefont {L.}~\bibnamefont {Heaney}},
  \bibinfo {author} {\bibfnamefont {M.~C.~O.}\ \bibnamefont {Aguiar}},\ and\
  \bibinfo {author} {\bibfnamefont {M.~F.}\ \bibnamefont {Santos}},\ }\bibfield
   {title} {\bibinfo {title} {Equilibrium and disorder-induced behavior in
  quantum light–matter systems},\ }\href
  {https://doi.org/10.1088/1367-2630/14/4/043033} {\bibfield  {journal}
  {\bibinfo  {journal} {New Journal of Physics}\ }\textbf {\bibinfo {volume}
  {14}},\ \bibinfo {pages} {043033} (\bibinfo {year} {2012})}\BibitemShut
  {NoStop}%
\bibitem [{\citenamefont {Ghoshal}\ \emph {et~al.}(2020)\citenamefont
  {Ghoshal}, \citenamefont {Das}, \citenamefont {Sen(De)},\ and\ \citenamefont
  {Sen}}]{PhysRevA.101.053805}%
  \BibitemOpen
  \bibfield  {author} {\bibinfo {author} {\bibfnamefont {A.}~\bibnamefont
  {Ghoshal}}, \bibinfo {author} {\bibfnamefont {S.}~\bibnamefont {Das}},
  \bibinfo {author} {\bibfnamefont {A.}~\bibnamefont {Sen(De)}},\ and\ \bibinfo
  {author} {\bibfnamefont {U.}~\bibnamefont {Sen}},\ }\bibfield  {title}
  {\bibinfo {title} {Population inversion and entanglement in single and double
  glassy jaynes-cummings models},\ }\href
  {https://doi.org/10.1103/PhysRevA.101.053805} {\bibfield  {journal} {\bibinfo
   {journal} {Phys. Rev. A}\ }\textbf {\bibinfo {volume} {101}},\ \bibinfo
  {pages} {053805} (\bibinfo {year} {2020})}\BibitemShut {NoStop}%
\bibitem [{\citenamefont {Ma}\ \emph {et~al.}(2022)\citenamefont {Ma},
  \citenamefont {Li},\ and\ \citenamefont {Tan}}]{PhysRevB.105.165432}%
  \BibitemOpen
  \bibfield  {author} {\bibinfo {author} {\bibfnamefont {J.-L.}\ \bibnamefont
  {Ma}}, \bibinfo {author} {\bibfnamefont {Q.}~\bibnamefont {Li}},\ and\
  \bibinfo {author} {\bibfnamefont {L.}~\bibnamefont {Tan}},\ }\bibfield
  {title} {\bibinfo {title} {Ergodic and nonergodic phases in a one-dimensional
  clean jaynes-cummings-hubbard system with detuning},\ }\href
  {https://doi.org/10.1103/PhysRevB.105.165432} {\bibfield  {journal} {\bibinfo
   {journal} {Phys. Rev. B}\ }\textbf {\bibinfo {volume} {105}},\ \bibinfo
  {pages} {165432} (\bibinfo {year} {2022})}\BibitemShut {NoStop}%
\bibitem [{\citenamefont {Oganesyan}\ and\ \citenamefont
  {Huse}(2007)}]{PhysRevB.75.155111}%
  \BibitemOpen
  \bibfield  {author} {\bibinfo {author} {\bibfnamefont {V.}~\bibnamefont
  {Oganesyan}}\ and\ \bibinfo {author} {\bibfnamefont {D.~A.}\ \bibnamefont
  {Huse}},\ }\bibfield  {title} {\bibinfo {title} {Localization of interacting
  fermions at high temperature},\ }\href
  {https://doi.org/10.1103/PhysRevB.75.155111} {\bibfield  {journal} {\bibinfo
  {journal} {Phys. Rev. B}\ }\textbf {\bibinfo {volume} {75}},\ \bibinfo
  {pages} {155111} (\bibinfo {year} {2007})}\BibitemShut {NoStop}%
\bibitem [{\citenamefont {Bianchi}\ \emph {et~al.}(2022)\citenamefont
  {Bianchi}, \citenamefont {Hackl}, \citenamefont {Kieburg}, \citenamefont
  {Rigol},\ and\ \citenamefont {Vidmar}}]{PRXQuantum.3.030201}%
  \BibitemOpen
  \bibfield  {author} {\bibinfo {author} {\bibfnamefont {E.}~\bibnamefont
  {Bianchi}}, \bibinfo {author} {\bibfnamefont {L.}~\bibnamefont {Hackl}},
  \bibinfo {author} {\bibfnamefont {M.}~\bibnamefont {Kieburg}}, \bibinfo
  {author} {\bibfnamefont {M.}~\bibnamefont {Rigol}},\ and\ \bibinfo {author}
  {\bibfnamefont {L.}~\bibnamefont {Vidmar}},\ }\bibfield  {title} {\bibinfo
  {title} {Volume-law entanglement entropy of typical pure quantum states},\
  }\href {https://doi.org/10.1103/PRXQuantum.3.030201} {\bibfield  {journal}
  {\bibinfo  {journal} {PRX Quantum}\ }\textbf {\bibinfo {volume} {3}},\
  \bibinfo {pages} {030201} (\bibinfo {year} {2022})}\BibitemShut {NoStop}%
\bibitem [{\citenamefont {Page}(1993)}]{PhysRevLett.71.1291}%
  \BibitemOpen
  \bibfield  {author} {\bibinfo {author} {\bibfnamefont {D.~N.}\ \bibnamefont
  {Page}},\ }\bibfield  {title} {\bibinfo {title} {Average entropy of a
  subsystem},\ }\href {https://doi.org/10.1103/PhysRevLett.71.1291} {\bibfield
  {journal} {\bibinfo  {journal} {Phys. Rev. Lett.}\ }\textbf {\bibinfo
  {volume} {71}},\ \bibinfo {pages} {1291} (\bibinfo {year}
  {1993})}\BibitemShut {NoStop}%
\bibitem [{\citenamefont {Khemani}\ \emph
  {et~al.}(2017{\natexlab{a}})\citenamefont {Khemani}, \citenamefont {Sheng},\
  and\ \citenamefont {Huse}}]{PhysRevLett.119.075702}%
  \BibitemOpen
  \bibfield  {author} {\bibinfo {author} {\bibfnamefont {V.}~\bibnamefont
  {Khemani}}, \bibinfo {author} {\bibfnamefont {D.~N.}\ \bibnamefont {Sheng}},\
  and\ \bibinfo {author} {\bibfnamefont {D.~A.}\ \bibnamefont {Huse}},\
  }\bibfield  {title} {\bibinfo {title} {Two universality classes for the
  many-body localization transition},\ }\href
  {https://doi.org/10.1103/PhysRevLett.119.075702} {\bibfield  {journal}
  {\bibinfo  {journal} {Phys. Rev. Lett.}\ }\textbf {\bibinfo {volume} {119}},\
  \bibinfo {pages} {075702} (\bibinfo {year} {2017}{\natexlab{a}})}\BibitemShut
  {NoStop}%
\bibitem [{\citenamefont {Khemani}\ \emph
  {et~al.}(2017{\natexlab{b}})\citenamefont {Khemani}, \citenamefont {Lim},
  \citenamefont {Sheng},\ and\ \citenamefont {Huse}}]{PhysRevX.7.021013}%
  \BibitemOpen
  \bibfield  {author} {\bibinfo {author} {\bibfnamefont {V.}~\bibnamefont
  {Khemani}}, \bibinfo {author} {\bibfnamefont {S.~P.}\ \bibnamefont {Lim}},
  \bibinfo {author} {\bibfnamefont {D.~N.}\ \bibnamefont {Sheng}},\ and\
  \bibinfo {author} {\bibfnamefont {D.~A.}\ \bibnamefont {Huse}},\ }\bibfield
  {title} {\bibinfo {title} {Critical properties of the many-body localization
  transition},\ }\href {https://doi.org/10.1103/PhysRevX.7.021013} {\bibfield
  {journal} {\bibinfo  {journal} {Phys. Rev. X}\ }\textbf {\bibinfo {volume}
  {7}},\ \bibinfo {pages} {021013} (\bibinfo {year}
  {2017}{\natexlab{b}})}\BibitemShut {NoStop}%
\bibitem [{\citenamefont {\ifmmode \check{Z}\else
  \v{Z}\fi{}nidari\ifmmode~\check{c}\else \v{c}\fi{}}\ \emph
  {et~al.}(2008)\citenamefont {\ifmmode \check{Z}\else
  \v{Z}\fi{}nidari\ifmmode~\check{c}\else \v{c}\fi{}}, \citenamefont {Prosen},\
  and\ \citenamefont {Prelov\ifmmode~\check{s}\else
  \v{s}\fi{}ek}}]{PhysRevB.77.064426}%
  \BibitemOpen
  \bibfield  {author} {\bibinfo {author} {\bibfnamefont {M.}~\bibnamefont
  {\ifmmode \check{Z}\else \v{Z}\fi{}nidari\ifmmode~\check{c}\else
  \v{c}\fi{}}}, \bibinfo {author} {\bibfnamefont {T.~c.~v.}\ \bibnamefont
  {Prosen}},\ and\ \bibinfo {author} {\bibfnamefont {P.}~\bibnamefont
  {Prelov\ifmmode~\check{s}\else \v{s}\fi{}ek}},\ }\bibfield  {title} {\bibinfo
  {title} {Many-body localization in the heisenberg $xxz$ magnet in a random
  field},\ }\href {https://doi.org/10.1103/PhysRevB.77.064426} {\bibfield
  {journal} {\bibinfo  {journal} {Phys. Rev. B}\ }\textbf {\bibinfo {volume}
  {77}},\ \bibinfo {pages} {064426} (\bibinfo {year} {2008})}\BibitemShut
  {NoStop}%
\bibitem [{\citenamefont {Bardarson}\ \emph {et~al.}(2012)\citenamefont
  {Bardarson}, \citenamefont {Pollmann},\ and\ \citenamefont
  {Moore}}]{PhysRevLett.109.017202}%
  \BibitemOpen
  \bibfield  {author} {\bibinfo {author} {\bibfnamefont {J.~H.}\ \bibnamefont
  {Bardarson}}, \bibinfo {author} {\bibfnamefont {F.}~\bibnamefont
  {Pollmann}},\ and\ \bibinfo {author} {\bibfnamefont {J.~E.}\ \bibnamefont
  {Moore}},\ }\bibfield  {title} {\bibinfo {title} {Unbounded growth of
  entanglement in models of many-body localization},\ }\href
  {https://doi.org/10.1103/PhysRevLett.109.017202} {\bibfield  {journal}
  {\bibinfo  {journal} {Phys. Rev. Lett.}\ }\textbf {\bibinfo {volume} {109}},\
  \bibinfo {pages} {017202} (\bibinfo {year} {2012})}\BibitemShut {NoStop}%
\bibitem [{\citenamefont {Zhao}\ \emph {et~al.}(2020)\citenamefont {Zhao},
  \citenamefont {Narayanan},\ and\ \citenamefont {Cho}}]{PhysRevB.102.094201}%
  \BibitemOpen
  \bibfield  {author} {\bibinfo {author} {\bibfnamefont {Y.}~\bibnamefont
  {Zhao}}, \bibinfo {author} {\bibfnamefont {R.}~\bibnamefont {Narayanan}},\
  and\ \bibinfo {author} {\bibfnamefont {J.}~\bibnamefont {Cho}},\ }\bibfield
  {title} {\bibinfo {title} {Signatures of many-body localization and
  metastability by weak perturbation},\ }\href
  {https://doi.org/10.1103/PhysRevB.102.094201} {\bibfield  {journal} {\bibinfo
   {journal} {Phys. Rev. B}\ }\textbf {\bibinfo {volume} {102}},\ \bibinfo
  {pages} {094201} (\bibinfo {year} {2020})}\BibitemShut {NoStop}%
\bibitem [{\citenamefont {Mori}\ \emph {et~al.}(2018)\citenamefont {Mori},
  \citenamefont {Ikeda}, \citenamefont {Kaminishi},\ and\ \citenamefont
  {Ueda}}]{Mori_2018}%
  \BibitemOpen
  \bibfield  {author} {\bibinfo {author} {\bibfnamefont {T.}~\bibnamefont
  {Mori}}, \bibinfo {author} {\bibfnamefont {T.~N.}\ \bibnamefont {Ikeda}},
  \bibinfo {author} {\bibfnamefont {E.}~\bibnamefont {Kaminishi}},\ and\
  \bibinfo {author} {\bibfnamefont {M.}~\bibnamefont {Ueda}},\ }\bibfield
  {title} {\bibinfo {title} {Thermalization and prethermalization in isolated
  quantum systems: a theoretical overview},\ }\href
  {https://doi.org/10.1088/1361-6455/aabcdf} {\bibfield  {journal} {\bibinfo
  {journal} {Journal of Physics B: Atomic, Molecular and Optical Physics}\
  }\textbf {\bibinfo {volume} {51}},\ \bibinfo {pages} {112001} (\bibinfo
  {year} {2018})}\BibitemShut {NoStop}%
\bibitem [{\citenamefont {Bertini}\ \emph {et~al.}(2015)\citenamefont
  {Bertini}, \citenamefont {Essler}, \citenamefont {Groha},\ and\ \citenamefont
  {Robinson}}]{PhysRevLett.115.180601}%
  \BibitemOpen
  \bibfield  {author} {\bibinfo {author} {\bibfnamefont {B.}~\bibnamefont
  {Bertini}}, \bibinfo {author} {\bibfnamefont {F.~H.~L.}\ \bibnamefont
  {Essler}}, \bibinfo {author} {\bibfnamefont {S.}~\bibnamefont {Groha}},\ and\
  \bibinfo {author} {\bibfnamefont {N.~J.}\ \bibnamefont {Robinson}},\
  }\bibfield  {title} {\bibinfo {title} {Prethermalization and thermalization
  in models with weak integrability breaking},\ }\href
  {https://doi.org/10.1103/PhysRevLett.115.180601} {\bibfield  {journal}
  {\bibinfo  {journal} {Phys. Rev. Lett.}\ }\textbf {\bibinfo {volume} {115}},\
  \bibinfo {pages} {180601} (\bibinfo {year} {2015})}\BibitemShut {NoStop}%
\bibitem [{\citenamefont {D'Alessio}\ \emph {et~al.}(2016)\citenamefont
  {D'Alessio}, \citenamefont {Kafri}, \citenamefont {Polkovnikov},\ and\
  \citenamefont {Rigol}}]{d2016quantum}%
  \BibitemOpen
  \bibfield  {author} {\bibinfo {author} {\bibfnamefont {L.}~\bibnamefont
  {D'Alessio}}, \bibinfo {author} {\bibfnamefont {Y.}~\bibnamefont {Kafri}},
  \bibinfo {author} {\bibfnamefont {A.}~\bibnamefont {Polkovnikov}},\ and\
  \bibinfo {author} {\bibfnamefont {M.}~\bibnamefont {Rigol}},\ }\bibfield
  {title} {\bibinfo {title} {From quantum chaos and eigenstate thermalization
  to statistical mechanics and thermodynamics},\ }\href
  {https://doi.org/10.1080/00018732.2016.1198134} {\bibfield  {journal}
  {\bibinfo  {journal} {Advances in Physics}\ }\textbf {\bibinfo {volume}
  {65}},\ \bibinfo {pages} {239} (\bibinfo {year} {2016})}\BibitemShut
  {NoStop}%
\bibitem [{\citenamefont {Srednicki}(1999)}]{Srednicki_1999}%
  \BibitemOpen
  \bibfield  {author} {\bibinfo {author} {\bibfnamefont {M.}~\bibnamefont
  {Srednicki}},\ }\bibfield  {title} {\bibinfo {title} {The approach to thermal
  equilibrium in quantized chaotic systems},\ }\href
  {https://doi.org/10.1088/0305-4470/32/7/007} {\bibfield  {journal} {\bibinfo
  {journal} {Journal of Physics A: Mathematical and General}\ }\textbf
  {\bibinfo {volume} {32}},\ \bibinfo {pages} {1163} (\bibinfo {year}
  {1999})}\BibitemShut {NoStop}%
\bibitem [{\citenamefont {Deutsch}(1991)}]{PhysRevA.43.2046}%
  \BibitemOpen
  \bibfield  {author} {\bibinfo {author} {\bibfnamefont {J.~M.}\ \bibnamefont
  {Deutsch}},\ }\bibfield  {title} {\bibinfo {title} {Quantum statistical
  mechanics in a closed system},\ }\href
  {https://doi.org/10.1103/PhysRevA.43.2046} {\bibfield  {journal} {\bibinfo
  {journal} {Phys. Rev. A}\ }\textbf {\bibinfo {volume} {43}},\ \bibinfo
  {pages} {2046} (\bibinfo {year} {1991})}\BibitemShut {NoStop}%
\bibitem [{\citenamefont {Srednicki}(1994)}]{PhysRevE.50.888}%
  \BibitemOpen
  \bibfield  {author} {\bibinfo {author} {\bibfnamefont {M.}~\bibnamefont
  {Srednicki}},\ }\bibfield  {title} {\bibinfo {title} {Chaos and quantum
  thermalization},\ }\href {https://doi.org/10.1103/PhysRevE.50.888} {\bibfield
   {journal} {\bibinfo  {journal} {Phys. Rev. E}\ }\textbf {\bibinfo {volume}
  {50}},\ \bibinfo {pages} {888} (\bibinfo {year} {1994})}\BibitemShut
  {NoStop}%
\bibitem [{\citenamefont {LeBlond}\ \emph {et~al.}(2019)\citenamefont
  {LeBlond}, \citenamefont {Mallayya}, \citenamefont {Vidmar},\ and\
  \citenamefont {Rigol}}]{PhysRevE.100.062134}%
  \BibitemOpen
  \bibfield  {author} {\bibinfo {author} {\bibfnamefont {T.}~\bibnamefont
  {LeBlond}}, \bibinfo {author} {\bibfnamefont {K.}~\bibnamefont {Mallayya}},
  \bibinfo {author} {\bibfnamefont {L.}~\bibnamefont {Vidmar}},\ and\ \bibinfo
  {author} {\bibfnamefont {M.}~\bibnamefont {Rigol}},\ }\bibfield  {title}
  {\bibinfo {title} {Entanglement and matrix elements of observables in
  interacting integrable systems},\ }\href
  {https://doi.org/10.1103/PhysRevE.100.062134} {\bibfield  {journal} {\bibinfo
   {journal} {Phys. Rev. E}\ }\textbf {\bibinfo {volume} {100}},\ \bibinfo
  {pages} {062134} (\bibinfo {year} {2019})}\BibitemShut {NoStop}%
\bibitem [{\citenamefont {Brenes}\ \emph
  {et~al.}(2020{\natexlab{a}})\citenamefont {Brenes}, \citenamefont {LeBlond},
  \citenamefont {Goold},\ and\ \citenamefont {Rigol}}]{PhysRevLett.125.070605}%
  \BibitemOpen
  \bibfield  {author} {\bibinfo {author} {\bibfnamefont {M.}~\bibnamefont
  {Brenes}}, \bibinfo {author} {\bibfnamefont {T.}~\bibnamefont {LeBlond}},
  \bibinfo {author} {\bibfnamefont {J.}~\bibnamefont {Goold}},\ and\ \bibinfo
  {author} {\bibfnamefont {M.}~\bibnamefont {Rigol}},\ }\bibfield  {title}
  {\bibinfo {title} {Eigenstate thermalization in a locally perturbed
  integrable system},\ }\href {https://doi.org/10.1103/PhysRevLett.125.070605}
  {\bibfield  {journal} {\bibinfo  {journal} {Phys. Rev. Lett.}\ }\textbf
  {\bibinfo {volume} {125}},\ \bibinfo {pages} {070605} (\bibinfo {year}
  {2020}{\natexlab{a}})}\BibitemShut {NoStop}%
\bibitem [{\citenamefont {Brenes}\ \emph
  {et~al.}(2020{\natexlab{b}})\citenamefont {Brenes}, \citenamefont {Goold},\
  and\ \citenamefont {Rigol}}]{PhysRevB.102.075127}%
  \BibitemOpen
  \bibfield  {author} {\bibinfo {author} {\bibfnamefont {M.}~\bibnamefont
  {Brenes}}, \bibinfo {author} {\bibfnamefont {J.}~\bibnamefont {Goold}},\ and\
  \bibinfo {author} {\bibfnamefont {M.}~\bibnamefont {Rigol}},\ }\bibfield
  {title} {\bibinfo {title} {Low-frequency behavior of off-diagonal matrix
  elements in the integrable xxz chain and in a locally perturbed
  quantum-chaotic xxz chain},\ }\href
  {https://doi.org/10.1103/PhysRevB.102.075127} {\bibfield  {journal} {\bibinfo
   {journal} {Phys. Rev. B}\ }\textbf {\bibinfo {volume} {102}},\ \bibinfo
  {pages} {075127} (\bibinfo {year} {2020}{\natexlab{b}})}\BibitemShut
  {NoStop}%
\bibitem [{\citenamefont {Luitz}\ and\ \citenamefont
  {Bar~Lev}(2016)}]{PhysRevLett.117.170404}%
  \BibitemOpen
  \bibfield  {author} {\bibinfo {author} {\bibfnamefont {D.~J.}\ \bibnamefont
  {Luitz}}\ and\ \bibinfo {author} {\bibfnamefont {Y.}~\bibnamefont
  {Bar~Lev}},\ }\bibfield  {title} {\bibinfo {title} {Anomalous thermalization
  in ergodic systems},\ }\href {https://doi.org/10.1103/PhysRevLett.117.170404}
  {\bibfield  {journal} {\bibinfo  {journal} {Phys. Rev. Lett.}\ }\textbf
  {\bibinfo {volume} {117}},\ \bibinfo {pages} {170404} (\bibinfo {year}
  {2016})}\BibitemShut {NoStop}%
\bibitem [{\citenamefont {Mallayya}\ and\ \citenamefont
  {Rigol}(2019)}]{PhysRevLett.123.240603}%
  \BibitemOpen
  \bibfield  {author} {\bibinfo {author} {\bibfnamefont {K.}~\bibnamefont
  {Mallayya}}\ and\ \bibinfo {author} {\bibfnamefont {M.}~\bibnamefont
  {Rigol}},\ }\bibfield  {title} {\bibinfo {title} {Heating rates in
  periodically driven strongly interacting quantum many-body systems},\ }\href
  {https://doi.org/10.1103/PhysRevLett.123.240603} {\bibfield  {journal}
  {\bibinfo  {journal} {Phys. Rev. Lett.}\ }\textbf {\bibinfo {volume} {123}},\
  \bibinfo {pages} {240603} (\bibinfo {year} {2019})}\BibitemShut {NoStop}%
\bibitem [{\citenamefont {Aravinda}\ \emph {et~al.}(2021)\citenamefont
  {Aravinda}, \citenamefont {Rather},\ and\ \citenamefont
  {Lakshminarayan}}]{PhysRevResearch.3.043034}%
  \BibitemOpen
  \bibfield  {author} {\bibinfo {author} {\bibfnamefont {S.}~\bibnamefont
  {Aravinda}}, \bibinfo {author} {\bibfnamefont {S.~A.}\ \bibnamefont
  {Rather}},\ and\ \bibinfo {author} {\bibfnamefont {A.}~\bibnamefont
  {Lakshminarayan}},\ }\bibfield  {title} {\bibinfo {title} {From dual-unitary
  to quantum bernoulli circuits: Role of the entangling power in constructing a
  quantum ergodic hierarchy},\ }\href
  {https://doi.org/10.1103/PhysRevResearch.3.043034} {\bibfield  {journal}
  {\bibinfo  {journal} {Phys. Rev. Res.}\ }\textbf {\bibinfo {volume} {3}},\
  \bibinfo {pages} {043034} (\bibinfo {year} {2021})}\BibitemShut {NoStop}%
\bibitem [{\citenamefont {LeBlond}\ and\ \citenamefont
  {Rigol}(2020)}]{PhysRevE.102.062113}%
  \BibitemOpen
  \bibfield  {author} {\bibinfo {author} {\bibfnamefont {T.}~\bibnamefont
  {LeBlond}}\ and\ \bibinfo {author} {\bibfnamefont {M.}~\bibnamefont
  {Rigol}},\ }\bibfield  {title} {\bibinfo {title} {Eigenstate thermalization
  for observables that break hamiltonian symmetries and its counterpart in
  interacting integrable systems},\ }\href
  {https://doi.org/10.1103/PhysRevE.102.062113} {\bibfield  {journal} {\bibinfo
   {journal} {Phys. Rev. E}\ }\textbf {\bibinfo {volume} {102}},\ \bibinfo
  {pages} {062113} (\bibinfo {year} {2020})}\BibitemShut {NoStop}%
\end{thebibliography}
%

\end{document}